\documentclass[review, 12pt, 3p]{elsarticle}
\usepackage{lineno,hyperref}
\modulolinenumbers[5]

\usepackage{graphicx}

\usepackage{amsfonts,amssymb}

\usepackage{amsmath}
\usepackage{caption}

\usepackage{float} 
\usepackage{subcaption}

\makeatletter
\def\ps@pprintTitle{%
	\let\@oddhead\@empty
	\let\@evenhead\@empty
	\let\@oddfoot\@empty
	\let\@evenfoot\@oddfoot
}
\makeatother

\begin{document}

% \linenumbers 

\begin{frontmatter}

\title{Microstructure reconstruction of 2D/3D random materials via diffusion-based deep generative models
}
% \tnotetext[mytitlenote]{Fully documented templates are available in the elsarticle package on \href{http://www.ctan.org/tex-archive/macros/latex/contrib/elsarticle}{CTAN}.}

% %% Group authors per affiliation:
% \address{Radarweg 29, Amsterdam}
% \fntext[myfootnote]{Since 1880.}

\author[mymainaddress]{Xianrui Lyu}

\author[mymainaddress]{Xiaodan Ren\corref{mycorrespondingauthor}}
\cortext[mycorrespondingauthor]{Corresponding author}
\ead{rxdtj@tongji.edu.cn}

\address[mymainaddress]{College of Civil Engineering, Tongji University, Shanghai 200092, PR China}
% \address[mysecondaryaddress]{360 Park Avenue South, New York}

\begin{abstract}
	Microstructure reconstruction serves as a crucial foundation for establishing Process-Structure-Property (PSP) relationship in material design. Confronting the limitations of variational autoencoder and generative adversarial network within generative modeling, this study adopted the denoising diffusion probability model (DDPM) to learn the probability distribution of high-dimensional raw data and successfully reconstructed the microstructures of various composite materials, such as inclusion materials, spinodal decomposition materials, chessboard materials, fractal noise materials, and so on. The quality of generated microstructure was evaluated using quantitative measures like spatial correlation functions and Fourier descriptor. On this basis, this study also successfully achieved the regulation of microstructure randomness and the generation of gradient materials through continuous interpolation in latent space using denoising diffusion implicit model (DDIM). Furthermore, the two-dimensional microstructure reconstruction is extended to three-dimensional framework and integrates permeability as a feature encoding embedding. This enables the conditional generation of three-dimensional microstructures for random porous materials within a defined permeability range. The permeabilities of these generated microstructures were further validated through the application of the Boltzmann method.
\end{abstract}


\end{frontmatter}

\section{Introduction}

% \paragraph{Installation} If the document class \emph{elsarticle} is not available on your computer, you can download and install the system package \emph{texlive-publishers} (Linux) or install the \LaTeX\ package \emph{elsarticle} using the package manager of your \TeX\ installation, which is typically \TeX\ Live or Mik\TeX.
The geometric morphologies of engineering materials play a pivotal role in elucidating their performance characteristics, such as the light-capturing efficiency in silicon solar cells \cite{RN683}, mass transport in porous materials \cite{RN638} and the electrical properties of polymer nanocomposite \cite{RN637}. Consequently, in the field of high-throughput computational materials science, there has been a notable shift away from traditional trial-and-error methods for material discovery. Instead, there is a growing emphasis on inverse design, focusing on the elucidation of the intricate relationships between processing, structure, and property (PSP) \cite{RN88}. The pursuit of identifying the mapping between PSP enables targeted microstructure design to guide the development of novel materials with desired performance characteristics, which has become a new research framework in the field of material development.

However, the emerging paradigm in materials science has ushered in fresh challenges. The primary challenges inherent in PSP-based inverse design involve extracting intrinsic metrics from intricate microstructures and achieving their equivalent reconstruction. In response to these challenges, scholars have conducted extensive research on microstructure characterization and reconstruction (MCR) technology to date \cite{RN382}. Specifically, they have commonly utilized statistical descriptors \cite{RN606}, including the two-point correlation function, linear correlation function, and two-point clustering correlation function, among others, as well as the spectral density function \cite{RN677, RN676}. These quantitative characterization descriptors are essentially statistical in nature, primarily focusing on the low-order statistical aspects of microstructure. For instance, both the two-point correlation function and spectral density function provide the same second-order statistical information. Therefore, it's important to note that they inevitably result in a loss of detailed geometric information pertaining to the microstructure. Correspondingly, reconstruction methods that rely on these descriptors, such as the Yeong-Torquato (YT) algorithm \cite{RN154} based on spatial functions or the parameterized random field based on spectral density functions, generate microstructure samples that are not a complete description of the distribution of the original data. Additionally, it has been noted by some scholars \cite{RN673} that in certain heterogeneous materials, even when the microstructure shares identical low-order statistical information (e.g., two-point correlation function), the morphology and macroscopic properties can exhibit significant variations. These highlight the critical importance of incorporating high-order statistical information in reconstruction methods, not only for accurate material identification but also for revealing the profound influence of microstructures on shaping macroscopic properties. To overcome the limitation of insufficient morphological information in the YT method, many studies have indicated \cite{RN169, RN155, RN168} that enhancing the accuracy of reconstruction can be achieved by incorporating multiple statistical functions into the energy function. For instance, constructing the energy function by amalgamating the two-point correlation function, linear correlation function, and two-point clustering correlation function has been shown to be effective. However, it is essential to recognize that the YT method fundamentally operates as an iterative optimization algorithm. This approach necessitates the continual exchange of pixel points or voxel blocks within the initial configuration, followed by a comparison of the quantitative representation differences between the intermediate image and the real image in each iteration to minimize the energy function. The computational cost of this method increases with the complexity of the energy function, and the quality of the reconstruction largely depends on the careful choice of the characterization function. To sum up, the pursuit of a fast and efficient reconstruction method with high order statistical information applicable to diverse material systems remains a direction in need of exploration and focused research efforts.

In essence, the pixel (2D) or voxel representation (3D) of microstructures can be conceptualized as a complex probability distribution governed by ultra-high-dimensional random variables. Deep learning, one of its primary objectives and functions, is to learn the manifold structures and probability distributions within data \cite{RN643, RN644}, and inherently possesses a formidable capability to manipulate such high-dimensional probability distributions. Consequently, with the advent of deep learning in recent years, there has been a growing interest in leveraging deep learning for the description of high-dimensional probabilities associated with microstructures \cite{RN640, RN642, RN685}. This method is essentially an unsupervised generation model, and its goal is to generate new sample generation models with the same distribution as the training set, and achieve mapping between manifolds through the nonlinear mapping ability of neural networks. More specifically, its core issue is probability density estimation, which is mainly divided into explicit probability density estimation and implicit probability density estimation \cite{2017arXiv170100160G}. Explicit probability density estimation models, such as PixelRNN or PixelCNN, employ the chain rule of probability to reframe image generation as a sequence generation task, representing it as a product of joint conditional probabilities. While successful applications of reconstructing microstructures have been achieved in some materials \cite{RN128}, it requires a predetermined order for pixel generation, and the generation speed is often excessively slow. Conversely, the approximate explicit probability estimation model-variational autoencoder (VAE) \cite{2013arXiv1312.6114K} and the implicit probability density estimation model-generative adversarial network (GAN) \cite{2014arXiv1406.2661G} are widely used in the field of microstructure reconstruction in the past decade. VAE transforms the challenge of probability density estimation into one of function approximation. It achieves this by employing maximum likelihood estimation to bring a mixed Gaussian distribution closer to the true underlying distribution. Scholars \cite{RN462, RN187} have successfully reconstructed sandstone materials, inclusion materials, metamaterials, etc. using VAE. However, VAE itself faces the challenge of the "maximum likelihood training paradigm", which can lead to the generation of blurry images \cite{RN649}. In contrast to VAE, GAN achieve Nash Equilibrium between generator and discriminator through adversarial training, resulting in superior generation performance. Many scholars \cite{RN183, RN650, RN190, RN640, RN189} have proposed various data-driven microstructure reconstruction algorithms based on GANs. These approaches have been successfully applied in reconstructing microstructures in materials like polycrystalline grains, perovskite, carbon fiber rods, and rocks. Nevertheless, GAN training can be unstable due to the adversarial nature of the loss functions \cite{2017arXiv170104862A, 2016arXiv161102163M}. Additionally, GANs are vulnerable to modal collapse, a situation in which they repetitively generate the same image. These limitations also impede the practical application of GANs.

Recently, the diffusion model has emerged as the frontrunner in the field of AI-generated content (AIGC), surpassing both VAE and GAN. The most advanced text-to-image model, such as OpenAI's DALL·E 2 and Google's Imagen, are built upon the diffusion model. The inspiration behind the diffusion model stems from non-equilibrium thermodynamics and can be concisely characterized as a layered VAE. In contrast to GANs, the diffusion model operates without the need for adversarial training, bringing with it added advantages in terms of training efficiency, scalability, and parallelism. In terms of generative capabilities, in addition to achieving state-of-the-art image quality, diffusion models exhibit a robust ability to encompass various patterns and generate diverse outputs. Beyond image generation, diffusion models have demonstrated significant potential in a wide array of fields, including computer vision \cite{2021arXiv211200390A, 2022arXiv220511423B, 2020arXiv200806520C, 2021arXiv210615282H}, natural language processing \cite{2022arXiv220514217L, 2021arXiv211206749S} , time series modeling \cite{2020arXiv200909761K, 2022arXiv220809399L}, multimodal modeling \cite{2022arXiv220406125R, 2021arXiv211210752R}, and more. 

Encouragingly, diffusion models have also shone brightly in the fields of material synthesis and structural reconstruction. However, it mainly focuses on biomaterials or medical imaging \cite{2023arXiv230308440L, 2022arXiv220814125W}, such as protein modeling and cell shape prediction \cite{2023arXiv230303543G, RN671, RN668}. There are only a few studies dedicated to examining the reconstruction of the microstructures of composite materials. For example, some scholars \cite{DURETH2023105608, doi:10.1080/15376494.2023.2198528, unknown} successfully reconstructed the microstructure of polycrystalline alloys, carbonates, ceramics, fiber composites, and other materials based on diffusion model, and have verified the statistical similarity between the generated microstructure and the original microstructure. It is worth noting that the above reconstruction confined to two-dimensional microstructures, and there is also no sufficient discussion on the morphological characteristics and the latent space of the diffusion model. In another study, Nikolaos and Sun \cite{VLASSIS2023116126} achieved the generation of microstructures with target performance by introducing context feature vectors. However, this research was confined to the mechanical MNIST dataset, featuring $28\times28$ pixel images. There is a dearth of evaluation for random materials characterized by higher resolution and more intricate microstructures. Vlassis et al. \cite{2023arXiv230604411V} utilizes VAE to reduce the dimensionality of 3D point cloud structures to a low dimensional latent space, and reconstructs the 3D structure of sand particles after training the diffusion model in the latent space. Nevertheless, this approach only generates individual particles and lacks the incorporation of multi-scale features in microstructure generation.

Therefore, this study proposes an end-to-end microstructure reconstruction method based on data-driven denoising diffusion probability diffusion model (DDPM) for heterogeneous engineering materials in two and three dimensions. Initially, the microstructure datasets were established for various composite materials, including regular inclusions, chessboard structures, spinodal decomposition materials, and random materials. The above microstructures with resolutions of $64\times64$ and $128\times128$ were reconstructed successfully by DDPM, and the statistical descriptors such as two-point correlation function and linear correlation function are used to evaluate the quality of generated microstructures. Meanwhile, Fourier descriptor also was used in this study to verify the morphological similarity between the both. On this basis, this study fully explores and utilizes the latent space of diffusion models through the deterministic generation of denoising diffusion implicit model (DDIM), achieving the regulation of the randomness of generated microstructures. Following this, the study extended the two-dimensional DDPM reconstruction method to encompass three-dimensional conditional generation, which was verified to generate three-dimensional random porous materials with a specific range of permeability.

\section{Introduction to diffusion models}
\subsection{Denoising diffusion probability model}
\begin{figure}[ht]
    \centering
    \includegraphics[width=0.9\textwidth]{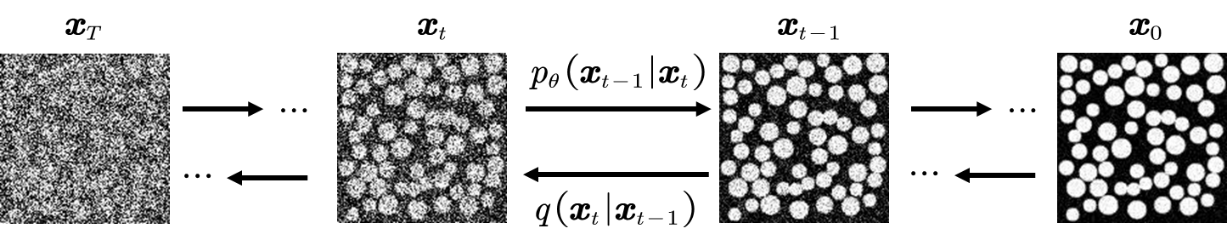}
    \caption{The forward noising process and reverse denoising process in diffusion model}
    \label{fig:label1}
    \end{figure}

The DDPM comprises two primary components: the forward diffusion process and the reverse diffusion process, analogous to the encoding and decoding phases in VAE. In the forward diffusion process, Gaussian noise is incrementally introduced to the original image tensor, ultimately transforming it into a noise image conforming to a standard normal distribution. Conversely, the reverse diffusion process entails the continuous removal of noise. A neural network, trained for this purpose, progressively eliminates noise from a given Gaussian noise image, ultimately restoring the original image from its noisy counterpart, as shown in Figure ~\ref{fig:label1}.

In contrast to the single-step mixed Gaussian distribution to approximate the original data distribution in VAE, the diffusion model employs a normal distribution to approximate incremental changes at each step. This approach enables the diffusion model to overcome the limitations typically associated with the fitting capacity of traditional single-step VAE.

Specifically, in the forward process, Gaussian noise $ \varepsilon $ is continuously added to the given initial data distribution $\mathbf{x}_0 \sim q(\mathbf{x})$, and the variance sequence of noise is $\beta _t$, which gradually increases with time steps. Each additional step of noise generates a new latent variable $\mathbf{x}_t$ with a distribution of $q\left( \mathbf{x}_{\mathbf{t}}\left| \mathbf{x}_{\mathbf{t}-1} \right. \right) $.
\begin{equation}
 q\left( \mathbf{x}_t|\mathbf{x}_{t-1} \right) =\mathcal{N}\left( \mathbf{x}_t;\sqrt{1-\beta _t}\mathbf{x}_{t-1},\beta _t\mathbf{I} \right) \quad q\left( \mathbf{x}_{1:T}|\mathbf{x}_0 \right) =\prod\limits_{t=1}^T{q}\left( \mathbf{x}_t|\mathbf{x}_{t-1} \right) 
\end{equation}

\begin{figure}[ht]
    \centering
    \includegraphics[width=0.9\textwidth]{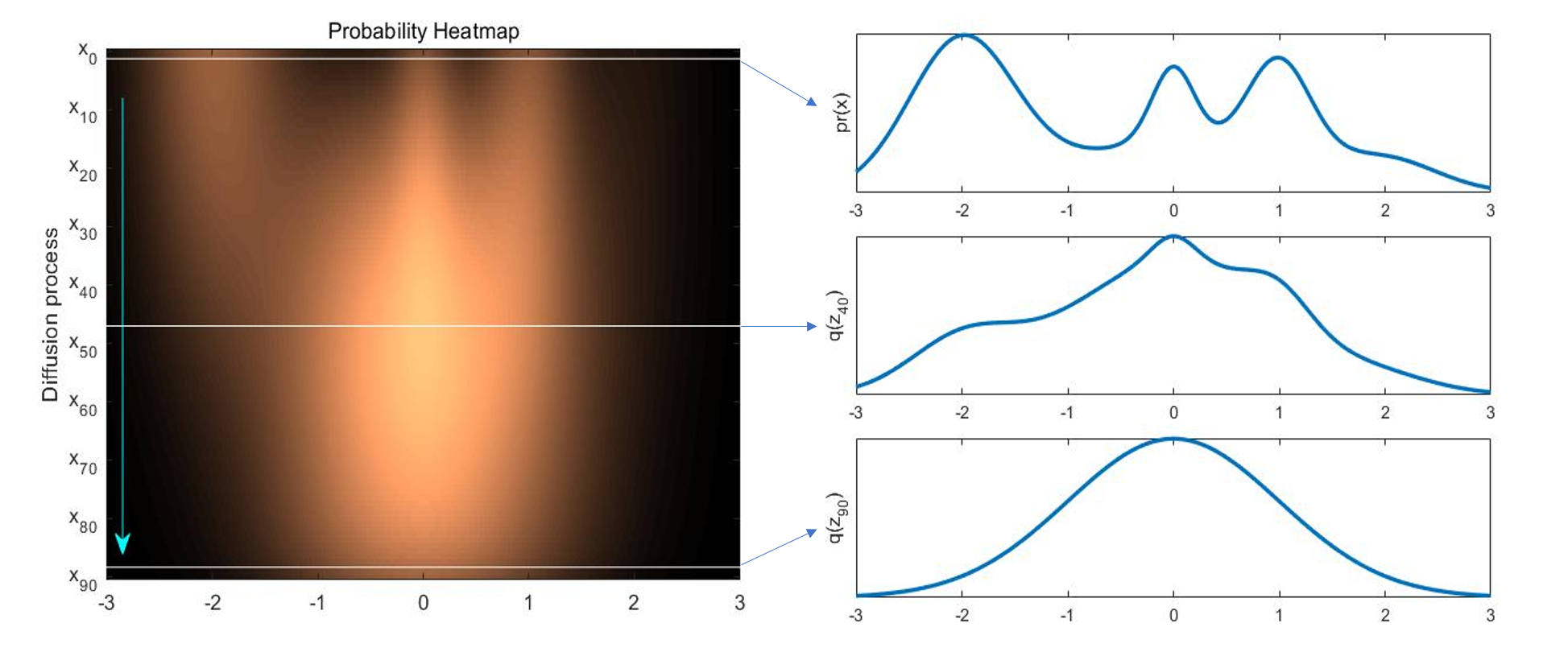}
    \caption{Probability density evolution in forward process}
    \label{fig:label16}
    \end{figure}

Based on the Markov chains, as $t$ progresses, the final data distribution $\mathbf{x}_T$  converges towards an anisotropic independent Gaussian distribution. The probability density evolution in the forward process is shown in Figure ~\ref{fig:label16}. Meanwhile, leveraging the characteristics of the Gaussian distribution, it can be inferred that at any given point during the forward process, it can be directly derived from $\mathbf{x}_0$ and $\beta _t$,
\begin{equation}
    q\left( \mathbf{x}_t|\mathbf{x}_0 \right) =\mathcal{N}\left( \mathbf{x}_t;\sqrt{\bar{\alpha}_t}\mathbf{x}_0,\left( 1-\bar{\alpha}_t \right) \mathbf{I} \right), 
\end{equation}
where $\alpha _t:=1-\beta _t$ and $\bar{\alpha}_t:=\prod_{i=0}^t{\alpha _i}$.

In the reverse process, it can be demonstrated that $q\left(\mathbf{x}_{t-1}|\mathbf{x}_t \right) $ also converges to a Gaussian distribution \cite{RN616}. Consequently, a parameterized distribution (a trainable neural network) $p_{\theta}\left( \mathbf{x}_{\mathbf{t}-1}\left| \mathbf{x}_{\mathbf{t}} \right. \right) $ is established to estimate the Gaussian distribution for the reverse diffusion, thereby,
\begin{equation}
 \quad p_{\theta}\left( \mathbf{x}_{t-1}|\mathbf{x}_t \right) =\mathcal{N}\left( \mathbf{x}_{t-1};\mathbf{\mu }_{\mathbf{\theta }}\left( \mathbf{x}_t,t \right) ,\Sigma _{\theta}\left( \mathbf{x}_t,t \right) \right).
\end{equation}
% \begin{equation}
% q\left( \mathbf{x}_{t-1}|\mathbf{x}_t,\mathbf{x}_0 \right) =q\left( \mathbf{x}_t|\mathbf{x}_{t-1},\mathbf{x}_0 \right) \frac{q\left( \mathbf{x}_{t-1}|\mathbf{x}_0 \right)}{q\left( \mathbf{x}_t|\mathbf{x}_0 \right)}\\
% \end{equation}

Combining the Bayesian formula with the forward diffusion formula Eq. 2, the posterior diffusion conditional probability can be explicitly expressed as:
\begin{footnotesize}
\begin{equation}
    \begin{split}
        q\left( \mathbf{x}_{t-1}|\mathbf{x}_t,\mathbf{x}_0 \right) &=q\left( \mathbf{x}_t|\mathbf{x}_{t-1},\mathbf{x}_0 \right) \frac{q\left( \mathbf{x}_{t-1}|\mathbf{x}_0 \right)}{q\left( \mathbf{x}_t|\mathbf{x}_0 \right)}\\
        &=\exp \left( -\frac{1}{2}\left( \left( \frac{\alpha _t}{\beta _t}+\frac{1}{1-\bar{\alpha}_{t-1}} \right) \mathbf{x}_{t-1}^{2}-\left( \frac{2\sqrt{\alpha _t}}{\beta _t}\mathbf{x}_t+\frac{2\sqrt{\alpha _t}}{1-\bar{\alpha}_t}\mathbf{x}_0 \right) \mathbf{x}_{t-1} +C\left( \mathbf{x}_t,\mathbf{x}_0 \right) \right) \right)\\. 
    \end{split}
\end{equation}
\end{footnotesize}

The mean and variance are,
\begin{equation}
    \begin{split}
    \tilde{\mu}_t\left( \mathbf{x}_t,\mathbf{x}_0 \right) 
    &=\left( \frac{\sqrt{\alpha _t}}{\beta _t}\mathbf{x}_t+\frac{\sqrt{\bar{\alpha}_t}}{1-\bar{\alpha}_t}\mathbf{x}_0 \right) /\left( \frac{\alpha _t}{\beta _t}+\frac{1}{1-\bar{\alpha}_{t-1}} \right)\\
    &=\frac{1}{\sqrt{\alpha _t}}\left( \mathbf{x}_t-\frac{1-\alpha _t}{\sqrt{1-\bar{\alpha}_t}}\varepsilon _t \right),
    \end{split}
\end{equation}
\begin{equation}
    \tilde{\beta}_t=1/\left( \frac{\alpha _t}{\beta _t}+\frac{1}{1-\bar{\alpha}_{t-1}} \right) =\frac{1-\bar{\alpha}_{t-1}}{1-\bar{\alpha}_t}\cdot \beta _t.
\end{equation}

It becomes clear that, under the conditions $\mathbf{x}_0$, the mean of the posterior conditional Gaussian distribution relies solely on noise distribution $\varepsilon _t$. Consequently, the parameterized distribution model merely requires the prediction of noise distributions for each time step to obtain the denoising map for each time interval during the reverse diffusion process. The detailed derivation of the above process can be referred to \cite{2020arXiv200611239H}.

\subsection{Denoising diffusion implicit model}
In DDPM, both forward and backward process are defined as a Markov chain. However, in the derivation of reverse diffusion process, only the edge distribution $q\left( \mathbf{x}_t\left| \mathbf{x}_0 \right. \right) $ is required. This implies that the inference distribution in the reverse process is not necessarily required to follow Markov chains. Therefore, in the DDIM, the distribution of reverse process is directly defined as \cite{2020arXiv201002502S},
\begin{equation}
	q_{\sigma}\left( \mathbf{x}_{t-1}|\mathbf{x}_t,\mathbf{x}_0 \right) =\mathcal{N}\left( \mathbf{x}_{t-1};\sqrt{\bar{\alpha}_{t-1}}\mathbf{x}_0+\sqrt{1-\bar{\alpha}_{t-1}-\sigma _{t}^{2}}\frac{\mathbf{x}_t-\sqrt{\bar{\alpha}_t}\mathbf{x}_0}{\sqrt{1-\bar{\alpha}_t}},\sigma _{t}^{2}\mathbf{I} \right).
\end{equation}

Among them, $\sigma _{t}^{2}$ is an uncertain real number, so $q_{\sigma}\left( \mathbf{x}_{t-1}|\mathbf{x}_t,\mathbf{x}_0 \right)$ is a series of inference distributions. It can be proven through mathematical induction that for all t,
\begin{equation}
	q_{\sigma}\left( \mathbf{x}_t|\mathbf{x}_0 \right) =\mathcal{N}\left( \mathbf{x}_t;\sqrt{\bar{\alpha}_t}\mathbf{x}_0,\left( 1-\bar{\alpha}_t \right) \mathbf{I} \right).
\end{equation}

By eliminating the Markov chain hypothesis in the reverse process, during the sampling, a larger step can be set to accelerate the generation process. According to Eq. 7, the following equation can be employed to facilitate sampling from $\mathbf{x}_t$  to $\mathbf{x}_{t-1}$ during the reverse diffusion process.
\begin{equation}
    \mathbf{x}_{t-1}=\sqrt{\alpha_{t-1}}\underbrace{\left(\frac{\mathbf{x}_t-\sqrt{1-\alpha_t}\epsilon_{\theta}(\mathbf{x}_t,t)}{\sqrt{\alpha_t}}\right)}_{\text{predicted }\mathbf{x_0}}+\underbrace{\sqrt{1-\alpha_{t-1}-\sigma_t^2}\cdot\epsilon_{\theta}(\mathbf{x}_t,t)}_{\text{direction pointing }to\quad\mathbf{x_t}}+\underbrace{\sigma_t\epsilon_t}_{\text{random}}
\end{equation}

The equation above decomposes the generation process into three components: the first is directly predicted $\mathbf{x}_0$ by $\mathbf{x}_t$, the second is the part pointing towards $\mathbf{x}_t$, and the third entails random noise $\epsilon _t$. The unknown parameter $\sigma _t$ is defined as follows:
\begin{equation}
    \sigma _{t}^{2}=\eta \cdot \tilde{\beta}_t=\eta \cdot \sqrt{\left( 1-\alpha _{t-1} \right) /\left( 1-\alpha _t \right)}\sqrt{\left( 1-\alpha _t/\alpha _{t-1} \right)}.
\end{equation}

When $\eta =1$, the variance setting in DDIM aligns with that in DDPM. Conversely, when $\eta =0$, random noise is absent, resulting in deterministic sampling. Specifically, once the initial random noise is established, DDIM's sample generation becomes a deterministic process. Considering that there is a continuity mapping between the latent space and the real data in DDIM, continuous sampling in the latent space can induce the dynamic evolution continuously of the real microstructure. This evolution also forms the foundation for subsequent regulation of randomness and the generation of gradient materials.

\subsection{Loss function}
Consistent with the fundamental objective of deep learning, the essence of DDPM resides in learning the underlying manifold structure from data and establishing parameter expressions. Building upon these objectives, it becomes crucial to minimize the Kullback-Leibler (KL) divergence between the parameterized probability distribution and the probability distribution of real images as effectively as possible. However, for the diffusion model, which operates across T time steps, it is necessary to extend the Evidence Lower Bound (ELBO) as used in VAE into a chain representation, which can be expressed as follows:
\begin{equation}
    \begin{aligned}
        \mathcal{L}_{VLB}& =\mathbb{E}_{q(x_{0:T})}\bigg[\log\frac{q(x_{1:T}\mid x_{0})}{p_{\theta}(x_{0:T})}\bigg]  \\
        &=\mathbb{E}_q\left[\underbrace{D_\mathrm{KL}\left(q(\mathbf{x}_T|\mathbf{x}_0)\parallel p_\theta(\mathbf{x}_T)\right)}_{L_T}+\sum_{t=2}^T\underbrace{D_\mathrm{KL}(q(\mathbf{x}_{t-1}|\mathbf{x}_t,\mathbf{x}_0)\parallel p_\theta(\mathbf{x}_{t-1}|\mathbf{x}_t)}_{L_{t-1}}-\underbrace{\log p_\theta(\mathbf{x}_0|\mathbf{x}_1)}_{L_0}\right].
        \end{aligned}
\end{equation}

The first item is independent of trainable parameters $\theta$ , and the ultimate objective is to optimize the KL divergence between two Gaussian distributions, $q\left( \mathbf{x}_{t-1}\left| \mathbf{x}_{t},\mathbf{x}_0 \right. \right) $ and $p_{\theta}\left( \mathbf{x}_{t-1}\left| \mathbf{x}_{t} \right. \right)$. Due to the existence of a closed form solution for the KL divergence of the multivariate Gaussian distribution, thus,
\begin{equation}
    \begin{aligned}
        \begin{split}
                L_{t-1}&=\mathbb{E}_{x_0,\epsilon}\left[ \frac{1}{2\lVert \Sigma _{\theta}\left( x_t,t \right) \rVert _{2}^{2}}|\bar{\mu}_t\left( x_t,x_0 \right) -\mu _{\theta}\left( x_t,t \right) |^2 \right]\\
                &=\mathbb{E}_{x_0,\epsilon}\left[ \frac{1}{2\lVert \Sigma _{\theta} \rVert _{2}^{2}}\lVert \frac{1}{\sqrt{\alpha _t}}\left( x_t-\frac{1-\alpha _t}{\sqrt{1-\bar{\alpha}_t}}\epsilon _t \right) -\frac{1}{\sqrt{\alpha _t}}\left( x_t-\frac{1-\alpha _t}{\sqrt{1-\bar{\alpha}_t}}\epsilon _{\theta}\left( x_t,t \right) \right) \rVert ^2 \right]\\
                &=\mathbb{E}_{x_0,\epsilon}\left[ \frac{\left( 1-\alpha _t \right) ^2}{2\alpha _t\left( 1-\bar{\alpha}_t \right) \lVert \Sigma _{\theta} \rVert _{2}^{2}}\lVert \epsilon _t-\epsilon _{\theta}\left( x_t,t \right) \rVert ^2 \right]\\
                &=\mathbb{E}_{x_0,\epsilon}\left[ \frac{\left( 1-\alpha _t \right) ^2}{2\alpha _t\left( 1-\bar{\alpha}_t \right) \lVert \Sigma _{\theta} \rVert _{2}^{2}}\lVert \epsilon _t-\epsilon _{\theta}\left( \sqrt{\bar{\alpha}_t}x_0+\sqrt{1-\bar{\alpha}_t}\epsilon _t,t \right) \rVert ^2 \right]\\.
        \end{split}
    \end{aligned}
\end{equation}

The final loss function can be simplified as,
\begin{equation}
    L_{t}^{\text{simple}}=\mathbb{E}_{x_0,\epsilon _t}\left[ \lVert \epsilon _t-\epsilon _{\theta}\left( \sqrt{\bar{\alpha}_t}x_0+\sqrt{1-\bar{\alpha}_t}\epsilon _t,t \right) \rVert ^2 \right].
\end{equation}

Indeed, this observation corroborates the conclusion made in the section 2.1. During the forward process, Gaussian noise is continuously injected into the clean image. Conversely, in the reverse process, neural networks are utilized to learn and progressively remove the noise. This iterative denoising process ultimately yields an image that adheres to the authentic data distribution.

\subsection{Conditional diffusion model}
The previously introduced DDPM and DDIM are essentially unconditional generators capable of producing microstructures resembling those in the original dataset in terms of morphology. However, in the realm of materials inverse design, the objective is to generate microstructures with specific functions or properties. As such, this study also introduces a conditional diffusion model designed to generate microstructures with predefined physical properties.

In the literature pertaining to diffusion models \cite{2021arXiv210505233D, 10030365}, Classifier Guidance has demonstrated efficacy in employing explicit classifiers to steer conditional generation. Nevertheless, this approach requires additional training of the image classifier which quality also can impact the effectiveness of conditional generation. Moreover, updating images through gradients can lead to adversarial attack effects, where generated images may deceive classifiers via imperceptible details.

Hence, in this study, we adopt the Classifier-Free Guidance scheme introduced by Google in 2022 to circumvent the aforementioned issues \cite{2022arXiv220712598H}. The crux of Classifier-Free Guidance involves substituting the explicit classifier with an implicit one, eliminating the need for direct calculations of the explicit classifier and its gradients.

More specifically, in addition to time feature encoding, this study incorporated supplementary feature encoding embeddings into the diffusion model, enabling control over various physical attributes, as shown in Figure ~\ref{fig:label3} and Figure ~\ref{fig:label4}. To leverage feature vectors for conditional generation, this study designate the reverse diffusion process, as defined in Eq.3, as a corresponding conditional probability distribution.
\begin{equation}
    p\left( \boldsymbol{x}_{t-1}|\boldsymbol{x}_t,\boldsymbol{y} \right) =\mathcal{N}\left( \boldsymbol{x}_{t-1};\boldsymbol{\mu }\left( \boldsymbol{x}_t,\boldsymbol{y} \right) ,\sigma _{t}^{2}\boldsymbol{I} \right) 
\end{equation}

The parameterization of the mean is expressed as,
\begin{equation}
    \boldsymbol{\mu }\left( \boldsymbol{x}_t,\boldsymbol{y} \right) =\frac{1}{\alpha _t}\left( \boldsymbol{x}_t-\frac{\beta _{t}^{2}}{\bar{\beta}_t}\boldsymbol{\epsilon }_{\boldsymbol{\theta }}\left( \boldsymbol{x}_t,\boldsymbol{y}, t \right) \right).
\end{equation}

The loss function for training is,
\begin{equation}
    \mathbb{E}_{\boldsymbol{x}_0,\boldsymbol{y}\sim \tilde{p}\left( \boldsymbol{x}_0,\boldsymbol{y} \right) ,\boldsymbol{\varepsilon }\sim \mathcal{N}\left( 0,\boldsymbol{I} \right)}\left[ \lVert \boldsymbol{\varepsilon }-\boldsymbol{\epsilon }_{\boldsymbol{\theta }}\left( \bar{\alpha}_t\boldsymbol{x}_0+\bar{\beta}_t\boldsymbol{\varepsilon}, \boldsymbol{y}, t \right) \rVert ^2 \right].
\end{equation}

Here, $\boldsymbol{y}$ represents the feature embedding corresponding to the physical attribute, effectively serving as the label for microscopic physical attributes, such as elastic modulus or permeability. Consequently, the conditional generation described above essentially constitutes a form of supervised learning.

\section{Introduction to network topology}
\subsection{3D U-Net}
\begin{figure}[ht]
    \centering
    \includegraphics[width=0.95\textwidth]{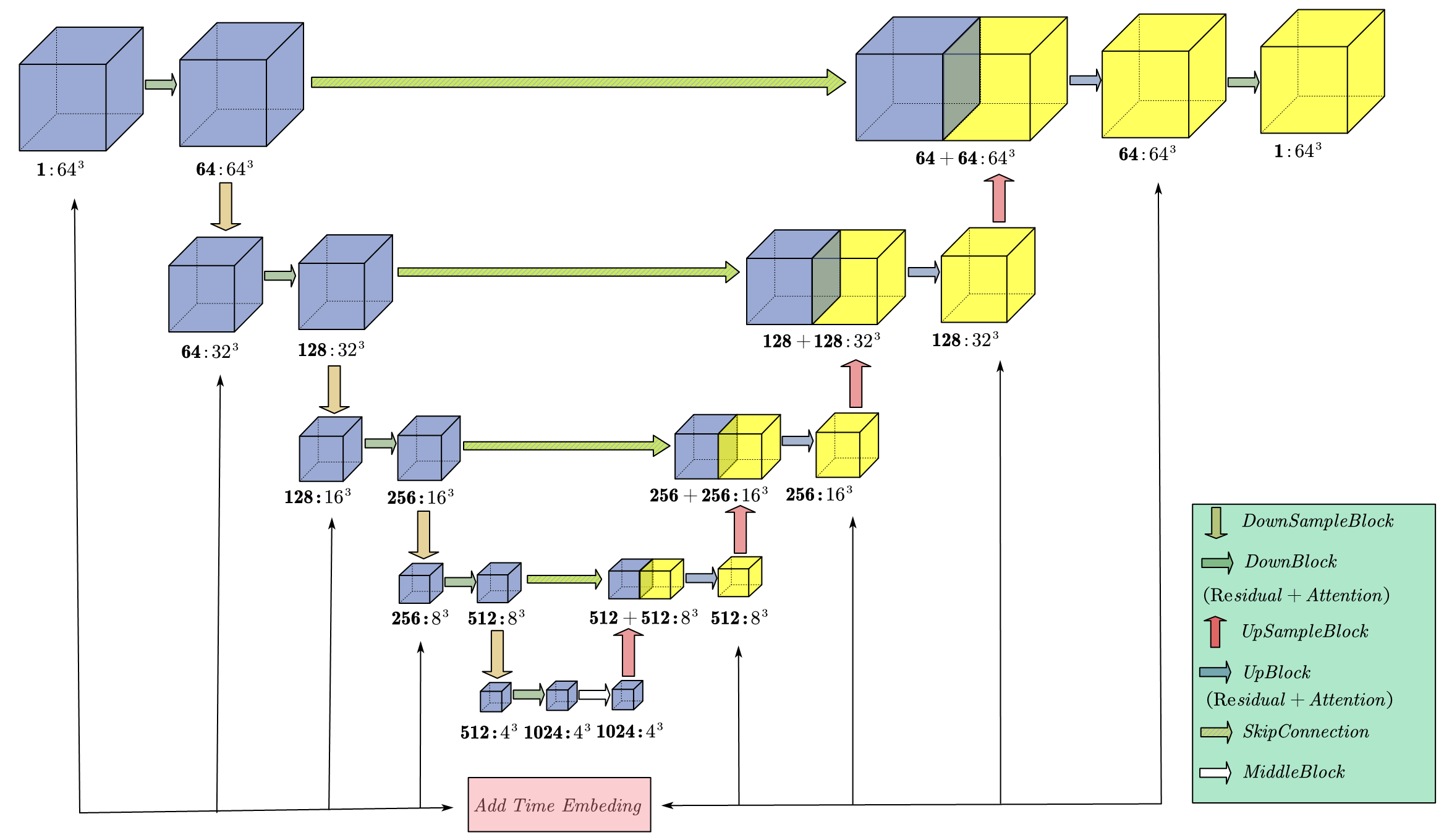}
    \caption{3D U-Net network architecture in diffusion model}
    \label{fig:label2}
    \end{figure}

Neural networks in DDPM and DDIM are tasked with taking noisy images as input and producing predicted noise, both of which have the same size as the image. Given these requirements, the primary network architecture employed in diffusion model is the U-Net network. In addition to making corresponding adjustments to binary microstructure images, this study further extends the two-dimensional diffusion model into three-dimensional framework. As a result, this section primarily centers on the introduction of the proposed three-dimensional U-Net network, as shown in Figure ~\ref{fig:label2}.

Specifically, U-Net adopts an encoder $\mathcal{D}\in \mathcal{N}\left( \boldsymbol{x;\theta } \right) =\mathcal{N}_{\mathcal{D}}\left( \boldsymbol{I;\theta }_{\mathcal{D}} \right) $ - decoder $\mathcal{U}\in \mathcal{N}\left( \boldsymbol{x;\theta } \right) =\mathcal{N}_{\mathcal{U}}\left( \boldsymbol{z;\theta }_{\mathcal{U}} \right) $ structure, with the network's architecture resembling the shape of the letter $'U'$. The encoder maps the image to the feature space $\boldsymbol{z}$, while the decoder maps the representation in the feature space back to the image space $\boldsymbol{I}$. Additionally, U-Net incorporates skip connection, connecting the output of the encoder to the corresponding decoder, facilitating feature fusion and enhancing gradient flow.
\begin{equation}
    I_{in}\subset \mathbb{R}^{D_I}\mapsto \mathcal{N}_{\mathcal{D}}\left( \boldsymbol{I;\theta }_{\mathcal{D}} \right) \mapsto Z\subset \mathbb{R}^{D_Z}\mapsto \mathcal{N}_{\mathcal{U}}\left( \boldsymbol{z;\theta }_{\mathcal{U}} \right) \mapsto I_{out}\subset \mathbb{R}^{D_I}
\end{equation}

In the implementation of U-Net in this study, both the encoder and decoder consist of five repeated modules. During the downsampling, the model's channel count changes sequentially as [1C, 2C, 4C, 8C, 16C], where $'C'$ is set to image size. The downsampling incorporates residual blocks and attention blocks. Each residual block comprises two convolutional blocks, each containing a normalization layer, a SiLU activation function, and a 3D convolutional layer with a stride of 2. The convolutional kernel size is $3\times3\times3$, with a padding of 1. The input undergoes feature extraction through the first convolutional block in residual block. Subsequently, time information is embedded into the feature map through time step embedding. Further feature extraction is then performed through the second convolutional block in residual block. Finally, the output of the second convolutional block is added to the output of the shortcut connection of to obtain the output of the residual Block. This design enables the residual block to effectively leverage temporal information and construct deep-level models, particularly when handling time series data. Moreover, skip connections are utilized to combine input feature maps with output feature maps, introducing residual learning to expedite network training and enhance gradient propagation in deep networks. To bolster the network's capacity to capture vital information, attention heads with resolutions of $16\times16\times16$ and $8\times8\times8$ are used for the output of residual blocks.
\begin{figure}[ht]
    \centering
    \includegraphics[width=0.8\textwidth]{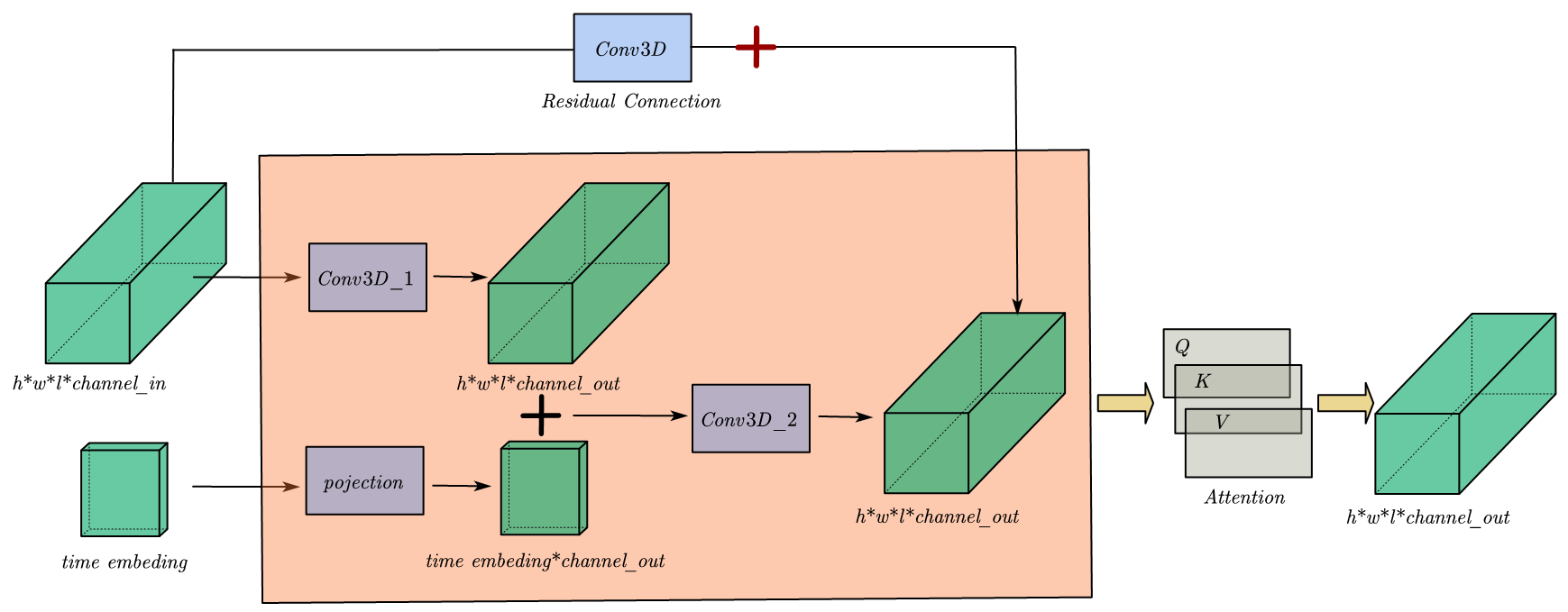}
    \caption{Schematic diagram of residual block structure}
    \label{fig:label15}
    \end{figure}

Similar to the encoder, the decoder consists of five repetitive modules, performing feature fusion and channel reduction through residual blocks and utilizing nearest neighbor interpolation to improve resolution. It's important to note that in these residual blocks, if the number of input channels does not match the number of output channels, a Conv3d is employed to adjust the number of channels in the input feature map to match the output feature map. If the input and output channel numbers are equal, identity mapping is directly used as the output of the shortcut connection.

\subsection{Training and sampling}
\begin{figure}[ht]
    \centering
    \includegraphics[width=0.8\textwidth]{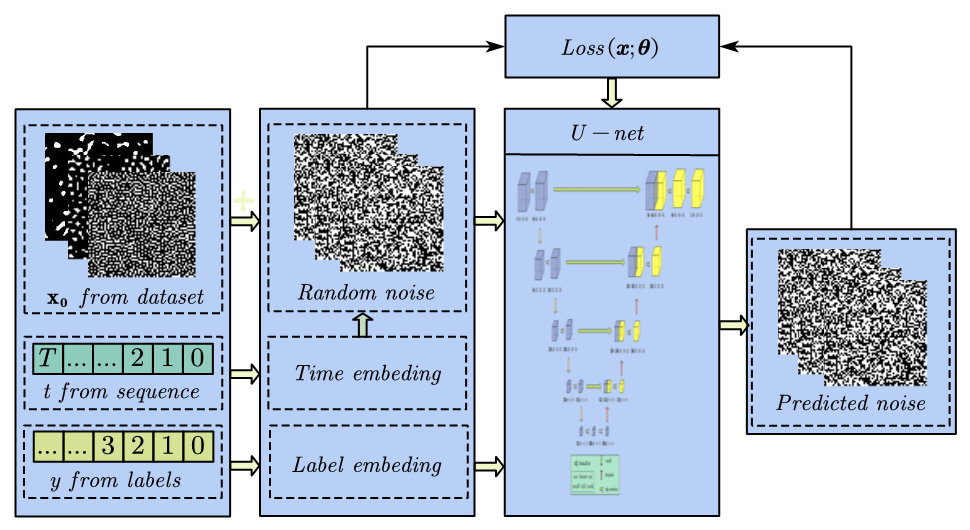}
    \caption{The training process in diffusion model}
    \label{fig:label3}
    \end{figure}

The training process, illustrated in the Figure ~\ref{fig:label3}, commences with the extraction of $\boldsymbol{x}_0$ from the training dataset. Following this, a random time step is selected from the time series. The noisy image $\boldsymbol{x}_t$ at this chosen time step is calculated using Eq. 2 and serves as the input for the model. Subsequently, the model generates the predicted noise at the corresponding time step. The loss function is defined as the mean square error (MSE) between the predicted noise and the noise introduced during the forward process at the same time step, as depicted in Eq. 14. The model in this study underwent 500 training iterations on computer with NVIDIA GeForce RTX 4070 Ti and was optimized using the Adam optimizer.

\begin{figure}[ht]
    \centering
    \includegraphics[width=0.8\textwidth]{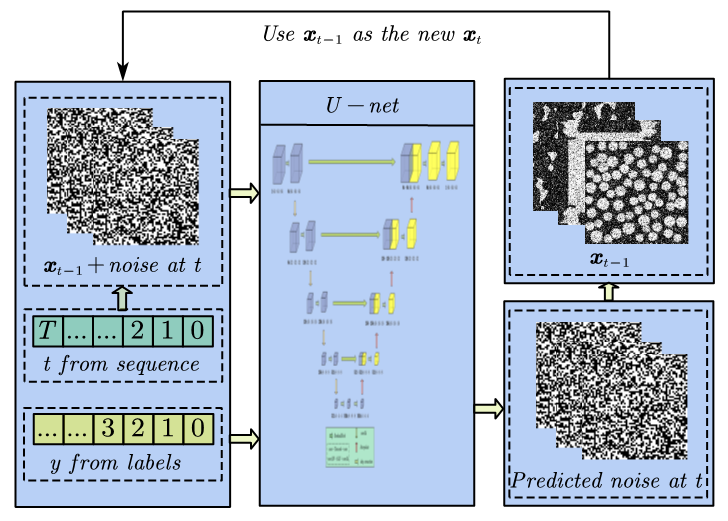}
    \caption{The sampling process in diffusion model}
    \label{fig:label4}
    \end{figure}

In the case of the trained model, when pure noise is provided as input, it initially generates the noise that was added during the forward process at time step 't'. Subsequently, it progressively denoises this noise to produce an image that conforms to the original data distribution. The detailed process is visually illustrated in the Figure ~\ref{fig:label4}.

\section{Evaluation criteria}
\subsection{Two-point correlation function}
The spatial correlation function, as an important descriptor for microstructure characterization, provides a reasonable quantitative description of the morphology of heterogeneous materials. Considering a two-phase random heterogeneous material, the definition of the indicator function is given as follows,
\begin{equation}
    \mathcal{I}^i\left( \boldsymbol{x} \right) =\left\{ \begin{array}{l}
        1,\ x\in V_i\\
        0,\ else\\
    \end{array} \right.,
\end{equation}
where $V_i\in \mathbb{R}^d$ is the space occupied by the phase $i$.

The statistical characterization of the spatial variation in two-phase materials, specifically the computation of the n-point correlation function, is described as follows \cite{RN590}:
\begin{equation}
    \begin{aligned}
        S_n^{(i)}(\boldsymbol{x}_1,\boldsymbol{x}_2,...,\boldsymbol{x}_n)& =\left<\mathcal{I}^{(i)}(\boldsymbol{x}_1)\mathcal{I}^{(i)}(\boldsymbol{x}_2)...\mathcal{I}^{(i)}(\boldsymbol{x}_n)\right>  \\
        &=P\left\{\mathcal{I}^{(i)}(\boldsymbol{x}_1)=1,\mathcal{I}^{(i)}(\boldsymbol{x}_2)=1,...,\mathcal{I}^{(i)}(\boldsymbol{x}_n)=1\right\}
        \end{aligned}
\end{equation}

The angle brackets $\left< \cdot \right> $ denote the average across all instances where $n$ points appear simultaneously in phase $i$, while $P\left\{ x_1,x_2,...,x_n \right\} $ represents the probability of $n$ points, denoted as $x_1,x_2,...,x_n$ randomly occurring at different positions within same phase $i$.

While the effectiveness of higher-order statistical descriptors in capturing collective microstructure phenomena is recognized, it's worth noting that their calculations can pose analytical and numerical challenges \cite{RN615}. Thus, the most commonly used is the two-point correlation function $S_2\left(r\right)$. For uniformly isotropic random materials, considering the translation and rotation invariance of the joint probability density distribution of the random field, the two-point correlation function disregards the position and direction information between points and is solely influenced by the relative distance between each point, $r^{ij}=x^j-x^i$. Therefore, particularly for statistically uniform and isotropic materials, the two-point correlation function can be further simplified as $S_{2}^{\left( i \right)}\left( \mathbf{x}_1,\mathbf{x}_2 \right) =S_{2}^{\left( i \right)}\left( |\mathbf{r|} \right)$ \cite{RN150}. It examines the correlation between the two endpoints of a vector within the same phase, and its form encapsulates crucial overarching characteristics of the microstructure \cite{RN590}.
\subsection{Linear correlation function}
Another method for the statistical description of microstructures is the linear path function, which serves as a statistical function employed to characterize aggregation within microstructures. From a probabilistic standpoint, it quantifies the probability of randomly placing a line on a microstructure image in such a way that the entire line falls within phase $i$. Specifically, it measures the extent of clustering along straight lines within a microstructure \cite{RN382}. This function is defined as a lower-order descriptor derived from a more complex fundamental function $\lambda _i$ which can capture specific information about phase connectivity and places greater emphasis on short-range correlation \cite{RN634}. The basic function $\lambda _i$ is defined as follows:
\begin{equation}
    \lambda ^i\left( \boldsymbol{x}_1,\boldsymbol{x}_2 \right) =\begin{cases}
        1,&		\text{if\ }\boldsymbol{x}_1\boldsymbol{x}_2\subset V_i\\
        0,&		\text{otherwise}\\
    \end{cases}
\end{equation}

For uniformly isotropic random materials, the above functions can be further simplified as,
\begin{equation}
    L\left( \boldsymbol{x}_1,\boldsymbol{x}_2 \right) =L\left( r \right) =\left< \lambda \left( \boldsymbol{x}_1,\boldsymbol{x}_2 \right) \right>.
\end{equation}

\subsection{Fourier descriptor}
Material properties are influenced not only by the spatial correlation between different phases but also by the morphology of phase boundaries. Factors such as the rate of dissolution/deposition are intimately linked to attributes like specific surface area and curvature \cite{RN600}. In order to further quantify the similarity between the generated microstructure phase profile and the original material phase profile, this study proposes the Fourier descriptor. This descriptor involves utilizing the Fourier transform of object boundary information, shifting contour characteristics from the spatial domain into the frequency domain, and extracting frequency domain information as the image's feature vector.

The boundaries between different phases in microstructure can be represented as coordinate sequences, $s\left( k \right) =\left[ x\left( k \right) ,y\left( k \right) \right] $, $k=0,1,2...K-1$. In addition, each coordinate can be represented by a complex number,
\begin{equation}
    s\left( k \right) =x\left( k \right) +jy\left( k \right).
\end{equation}

Perform Discrete Fourier Transform (DFT) on a one-dimensional sequence $s\left( k \right)$.
\begin{equation}
    a\left( u \right) =\sum_{k=0}^{K-1}{s}\left( k \right) \text{e}^{-j2\pi uk/K}
\end{equation}
Where $u=0,1,2...K-1$. The complex coefficient $a\left(u \right)$ is called the Fourier descriptor of the boundary, which can be used to describe boundary shape features in the frequency domain.

\section{Dataset generation and model validation}
\subsection{Generation of datasets}
To assess the efficacy of diffusion models in microstructure reconstruction with varying levels of complexity and randomness, this study initially established a database comprising various types of composite materials. These included fiber inclusion materials, texture materials, random materials, spinodal decomposition materials, chessboard structures, and Voronoi structured materials. Each microstructure in the database has the size of $64\times64\times1000$ (image size=64). Additionally, to evaluate the reconstruction performance of the model at different resolutions, the database of circular inclusion materials and fractal noise materials with a resolution of $128\times128$ has also been established. The database size for these materials was also 1000 samples.

Various reconstruction algorithms were employed to generate microstructures in the database. These algorithms encompassed random field methods based on random harmonic functions and reconstruction methods grounded in physical descriptors. 

\subsection{Reconstruction of 2D random materials based on DDPM}
\begin{figure}[htbp]
	\centering
	\begin{subfigure}{0.2\linewidth}
		\centering
		\includegraphics[width=0.8\linewidth]{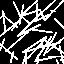}
		\caption{}
		\label{5.1}%文中引用该图片代号
	\end{subfigure}
	\centering
	\begin{subfigure}{0.2\linewidth}
		\centering
		\includegraphics[width=0.8\linewidth]{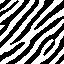}
		\caption{}
		\label{5.3}%文中引用该图片代号
	\end{subfigure}
	\centering
	\begin{subfigure}{0.2\linewidth}
		\centering
		\includegraphics[width=0.8\linewidth]{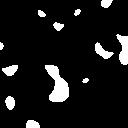}
		\caption{}
		\label{5.5}%文中引用该图片代号
	\end{subfigure}
    \centering
	\begin{subfigure}{0.2\linewidth}
		\centering
		\includegraphics[width=0.8\linewidth]{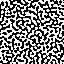}
		\caption{}
		\label{5.7}%文中引用该图片代号
	\end{subfigure}

    \centering
	\begin{subfigure}{0.2\linewidth}
		\centering
		\includegraphics[width=0.8\linewidth]{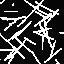}
		\caption{}
		\label{5.2}%文中引用该图片代号
	\end{subfigure}
    \centering
	\begin{subfigure}{0.2\linewidth}
		\centering
		\includegraphics[width=0.8\linewidth]{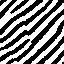}
		\caption{}
		\label{5.4}%文中引用该图片代号
	\end{subfigure}
    \centering
	\begin{subfigure}{0.2\linewidth}
		\centering
		\includegraphics[width=0.8\linewidth]{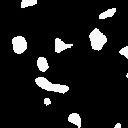}
		\caption{}
		\label{5.6}%文中引用该图片代号
	\end{subfigure}
    \centering
	\begin{subfigure}{0.2\linewidth}
		\centering
		\includegraphics[width=0.8\linewidth]{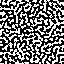}
		\caption{}
		\label{5.8}%文中引用该图片代号
	\end{subfigure}

    \centering
	\begin{subfigure}{0.2\linewidth}
		\centering
		\includegraphics[width=0.8\linewidth]{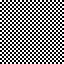}
		\caption{}
		\label{5.9}%文中引用该图片代号
	\end{subfigure}
    \centering
	\begin{subfigure}{0.2\linewidth}
		\centering
		\includegraphics[width=0.8\linewidth]{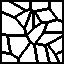}
		\caption{}
		\label{5.11}%文中引用该图片代号
	\end{subfigure}
    \centering
	\begin{subfigure}{0.2\linewidth}
		\centering
		\includegraphics[width=0.8\linewidth]{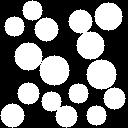}
		\caption{}
		\label{5.13}%文中引用该图片代号
	\end{subfigure}
    \centering
	\begin{subfigure}{0.2\linewidth}
		\centering
		\includegraphics[width=0.8\linewidth]{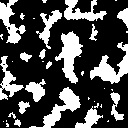}
		\caption{}
		\label{5.15}%文中引用该图片代号
	\end{subfigure}

    \centering
	\begin{subfigure}{0.2\linewidth}
		\centering
		\includegraphics[width=0.8\linewidth]{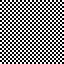}
		\caption{}
		\label{5.10}%文中引用该图片代号
	\end{subfigure}
    \centering
	\begin{subfigure}{0.2\linewidth}
		\centering
		\includegraphics[width=0.8\linewidth]{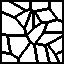}
		\caption{}
		\label{5.12}%文中引用该图片代号
	\end{subfigure}
    \centering
	\begin{subfigure}{0.2\linewidth}
		\centering
		\includegraphics[width=0.8\linewidth]{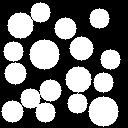}
		\caption{}
		\label{5.14}%文中引用该图片代号
	\end{subfigure}
    \centering
	\begin{subfigure}{0.2\linewidth}
		\centering
		\includegraphics[width=0.8\linewidth]{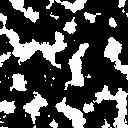}
		\caption{}
		\label{5.16}%文中引用该图片代号
	\end{subfigure}
	\caption{Comparison between microstructures generated based on DDPM and microstructures of raw materials;(a)-(d) and (i)-(l): Original microstructures; (e)-(h) and (m)-(p): Generated microstructures}
	\label{label5}
\end{figure}

In this section, the microstructures of random fiber inclusion materials, texture materials, random materials, spinodal decomposition materials, chessboard structures, and Voronoi structured materials, circular inclusion and fractal noise materials were successfully reconstructed based on DDPM. Figure ~\ref*{label5} provides a comparison between the original microstructure and the reconstructed microstructure, revealing no noticeable visual distinctions.

\begin{figure}[htbp]
	\centering
	\begin{subfigure}{0.45\linewidth}
		\centering
		\includegraphics[width=1.0\linewidth]{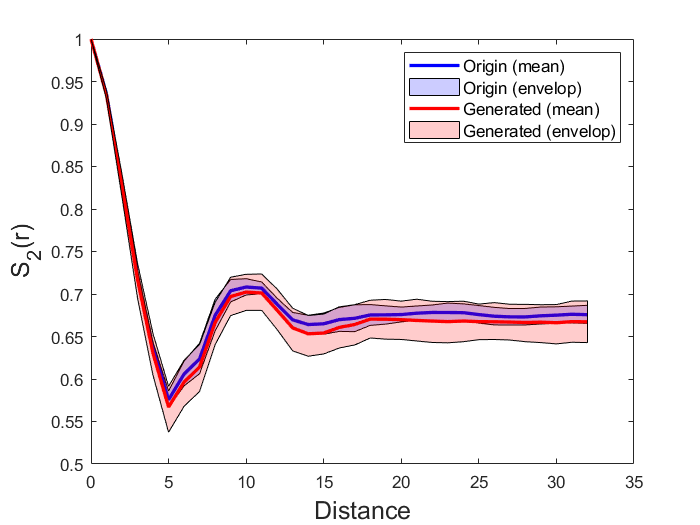}
		\caption{Texture material}
		\label{6.1}%文中引用该图片代号
	\end{subfigure}
	\centering
	\begin{subfigure}{0.45\linewidth}
		\centering
		\includegraphics[width=1.0\linewidth]{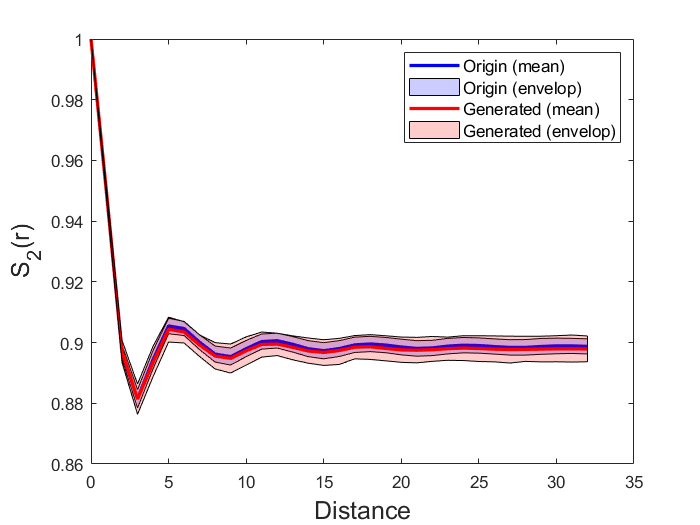}
		\caption{Spinodal decomposition}
		\label{6.2}%文中引用该图片代号
	\end{subfigure}

    \centering
	\begin{subfigure}{0.45\linewidth}
		\centering
		\includegraphics[width=1.0\linewidth]{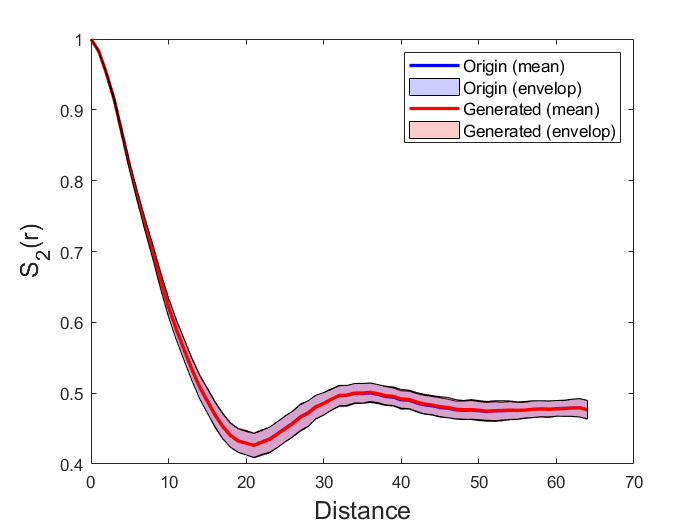}
		\caption{Circular inclusion}
		\label{6.3}%文中引用该图片代号
	\end{subfigure}
	\centering
	\begin{subfigure}{0.45\linewidth}
		\centering
		\includegraphics[width=1.0\linewidth]{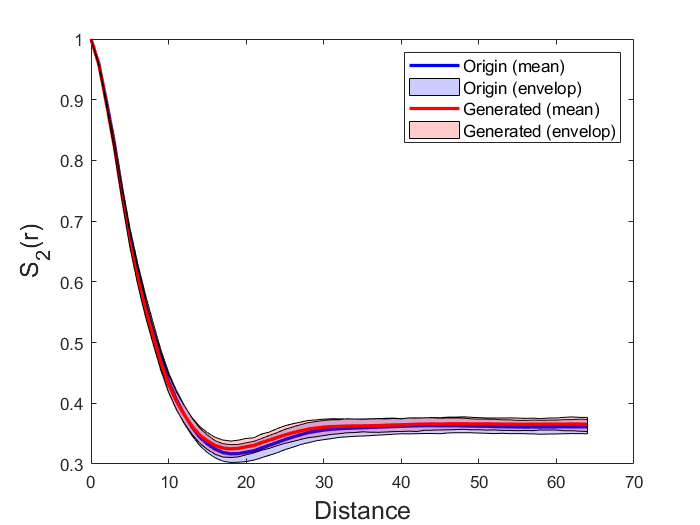}
		\caption{Fractal noise}
		\label{6.4}%文中引用该图片代号
	\end{subfigure}
    \caption{Comparison of $S_2 \left( r \right)$ between generated microstructure and original microstructure; (a):Texture material; (b): Spinodal decomposition; (c): Circular inclusion; (d): Fractal noise}
	\label{label6}
\end{figure}

\begin{figure}[htbp]
	\centering
	\begin{subfigure}{0.45\linewidth}
		\centering
		\includegraphics[width=1.0\linewidth]{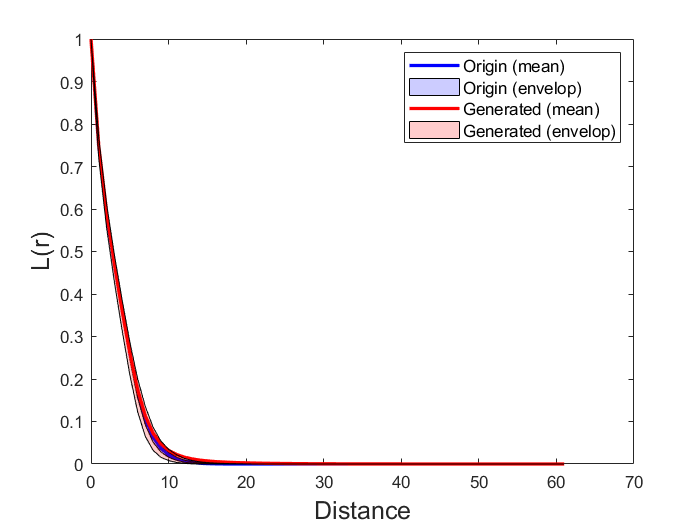}
		\caption{Texture material}
		\label{7.1}%文中引用该图片代号
	\end{subfigure}
	\centering
	\begin{subfigure}{0.45\linewidth}
		\centering
		\includegraphics[width=1.0\linewidth]{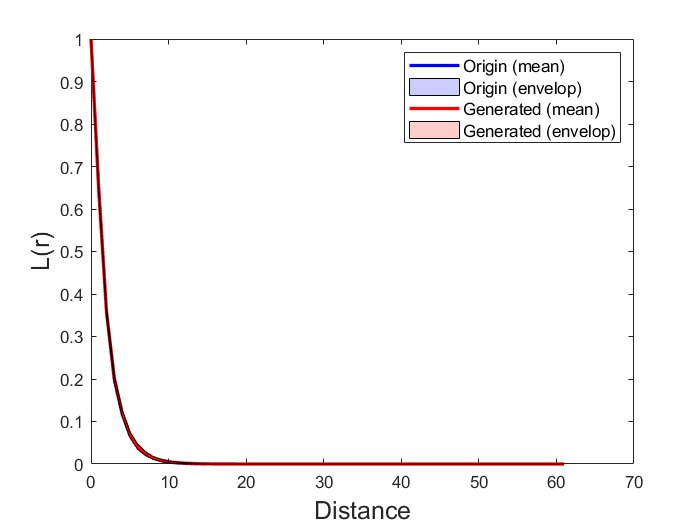}
		\caption{Spinodal decomposition}
		\label{7.2}%文中引用该图片代号
	\end{subfigure}

    \centering
	\begin{subfigure}{0.45\linewidth}
		\centering
		\includegraphics[width=1.0\linewidth]{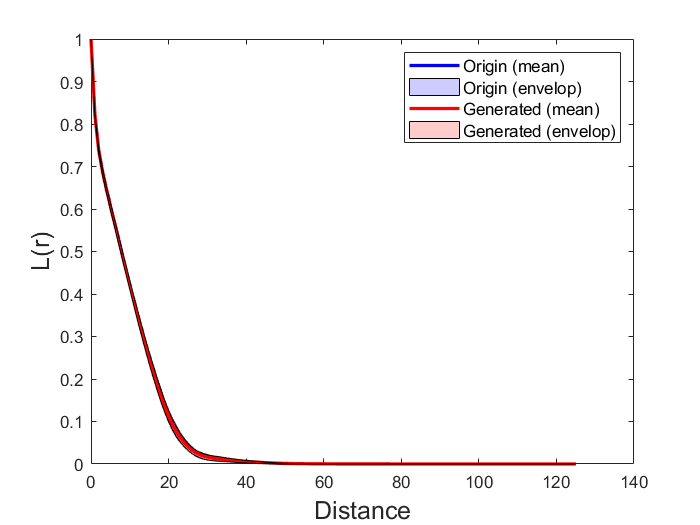}
		\caption{Circular inclusion}
		\label{7.3}%文中引用该图片代号
	\end{subfigure}
	\centering
	\begin{subfigure}{0.45\linewidth}
		\centering
		\includegraphics[width=1.0\linewidth]{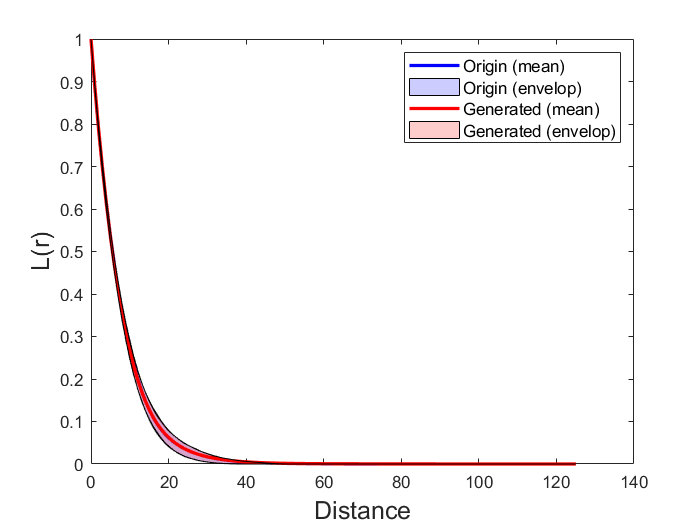}
		\caption{Fractal noise}
		\label{7.4}%文中引用该图片代号
	\end{subfigure}
    \caption{Comparison of $L\left(r\right)$ between generated microstructure and original microstructure; (a):Texture material; (b): Spinodal decomposition; (c): Circular inclusion; (d): Fractal noise}
	\label{label7}
\end{figure}

Following this, the spatial correlation functions, including the two-point correlation function $S_2\left( r \right)$ and linear path function $L\left(r\right)$, for various materials were computed to quantitatively assess the microstructure's morphology, as illustrated in the Figure ~\ref*{label6} and Figure ~\ref*{label7}. The comparison clearly demonstrates that the two-point correlation function and linear correlation function of both low and high-resolution microstructures reconstructed using DDPM exhibit a remarkable degree of consistency with the original material.

\begin{figure}[htbp]
	\centering
	\begin{subfigure}{0.45\linewidth}
		\centering
		\includegraphics[width=1.0\linewidth]{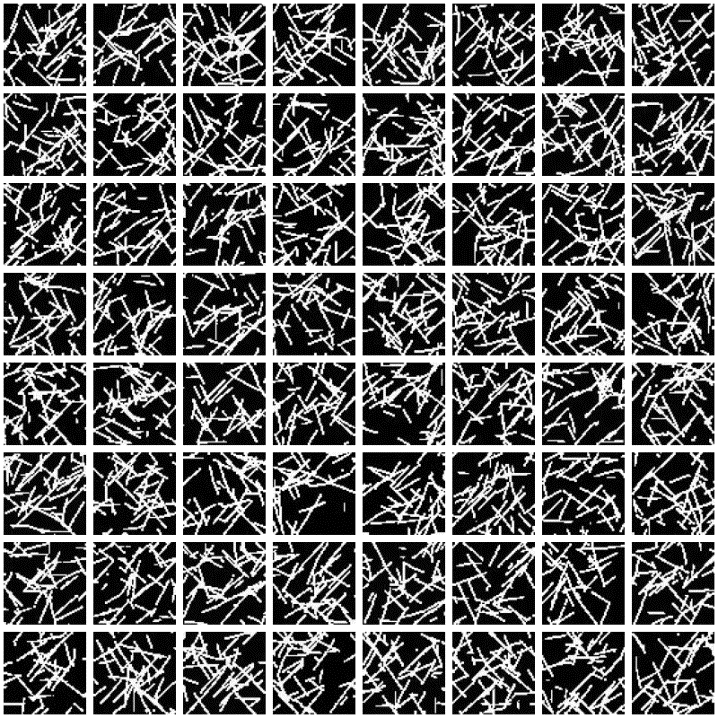}
		\caption{Fiber inclusions}
		\label{8.1}%文中引用该图片代号
	\end{subfigure}
	\centering
	\begin{subfigure}{0.45\linewidth}
		\centering
		\includegraphics[width=1.0\linewidth]{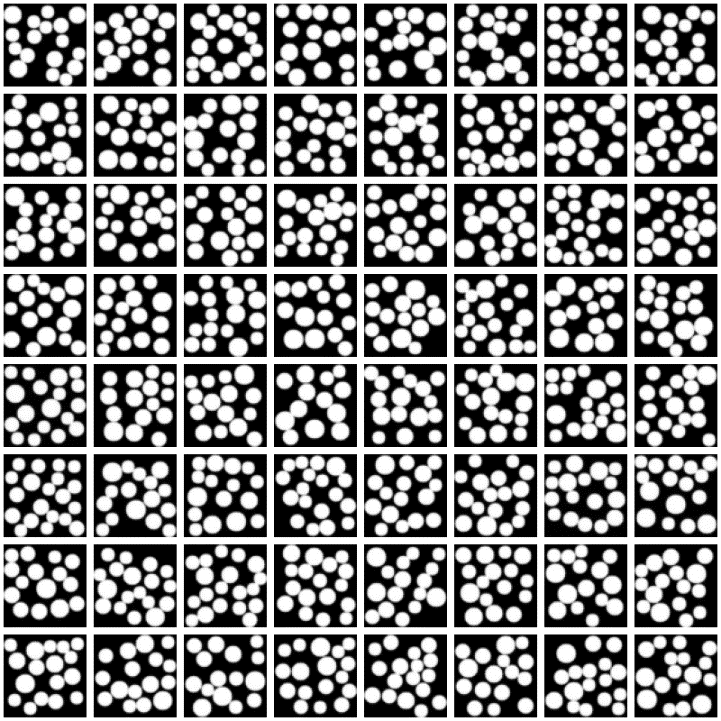}
		\caption{Circular inclusions}
		\label{8.2}%文中引用该图片代号
	\end{subfigure}

    \centering
	\begin{subfigure}{0.45\linewidth}
		\centering
		\includegraphics[width=1.0\linewidth]{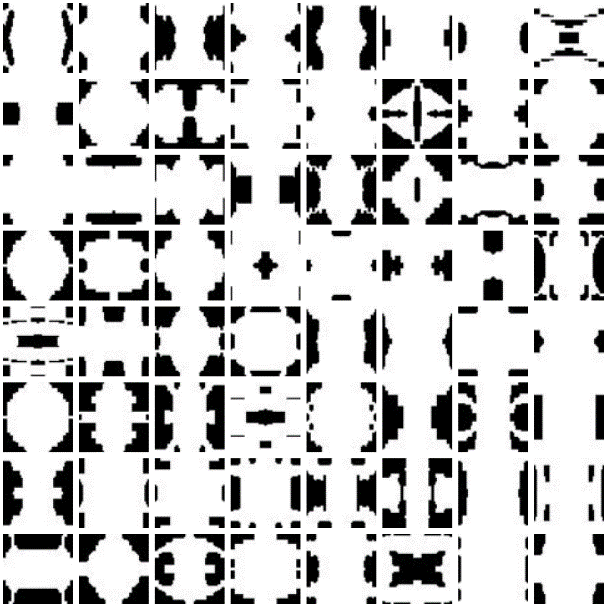}
		\caption{Metamaterials}
		\label{8.3}%文中引用该图片代号
	\end{subfigure}
	\centering
	\begin{subfigure}{0.45\linewidth}
		\centering
		\includegraphics[width=1.0\linewidth]{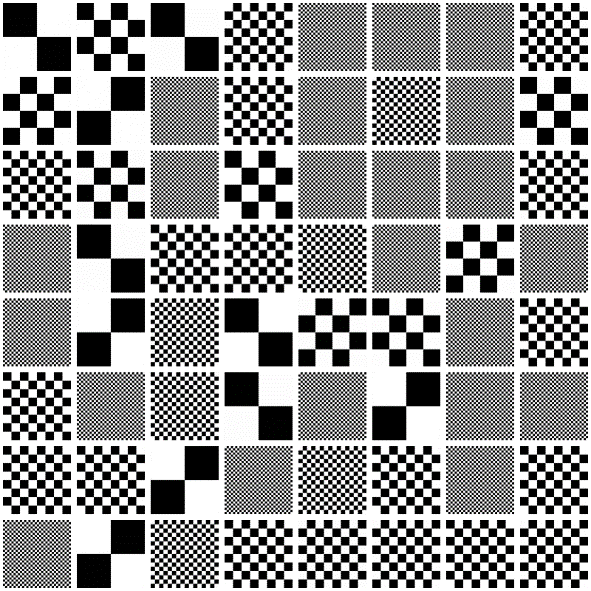}
		\caption{Chessboard structures}
		\label{8.4}%文中引用该图片代号
	\end{subfigure}
    \caption{Diversified generated microstructures based on diffusion model}
	\label{label8}
\end{figure}

To verify the diversity of DDPM, random sampling reconstruction was performed on fiber inclusions, circular inclusions, metamaterials and chessboard structures. On the other hand, this also suggests that DDPM has the capability to prevent the emergence of pattern collapse problems. The results are shown in Figure ~\ref*{label8}.

This study also employs Fourier descriptors to compare the original random microstructure with their reconstructed counterparts. As previously mentioned, Fourier descriptors effectively describe the shape of closed contours by transforming two-dimensional spatial boundary information into the frequency domain, where boundary shape information is represented by feature vectors of the Fourier descriptor (amplitude and phase).

\begin{figure}[htbp]
	\centering
	\begin{subfigure}{0.45\linewidth}
		\centering
		\includegraphics[width=1.0\linewidth]{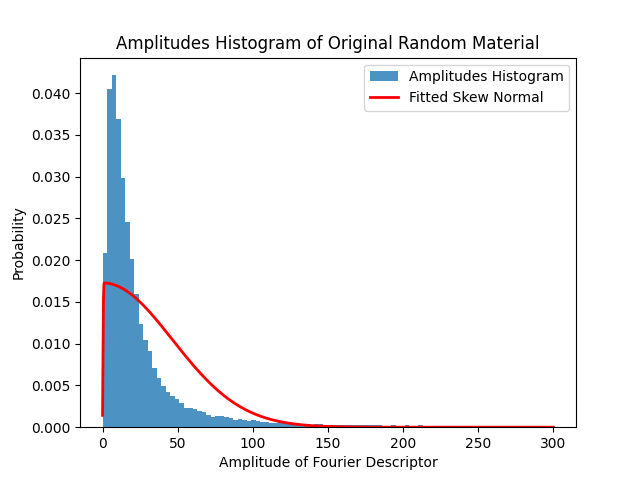}
		\caption{Original microstructure}
		\label{9.1}%文中引用该图片代号
	\end{subfigure}
	\centering
	\begin{subfigure}{0.45\linewidth}
		\centering
		\includegraphics[width=1.0\linewidth]{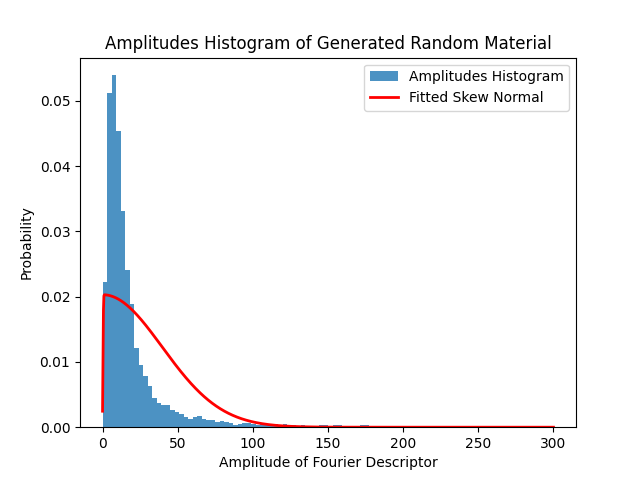}
		\caption{Reconstructed microstructure}
		\label{9.2}%文中引用该图片代号
	\end{subfigure}
    \caption{The feature distribution of Fourier descriptors}
	\label{label9}
\end{figure}

The study performs a statistical comparison by analyzing the histograms of the feature vectors of the Fourier descriptor for both the original and reconstructed random materials, as illustrated in the Figure ~\ref*{label9}. The figure reveals that the feature vectors of the reconstructed image's Fourier descriptor exhibit a similar clustering trend to those of the original image's Fourier descriptor, with their amplitudes primarily concentrated between 0 and 50.

To further quantify the similarity of boundary contours, a skewed distribution was employed to fit the histogram data. The skewness, mean shift, and standard deviation shift for the original microstructures were 182.78, 0.4, and 44.07, respectively. Meanwhile, for the reconstructed microstructures, these values were 155.19, 0.3, and 39.31, respectively. These results indicate that the reconstructed image effectively preserves shape-related information related to the overall contour or other features of the boundary.

\subsection{Reconstruction of 2D random materials based on DDIM}
As previously discussed, DDIM exhibits the capability to flexibly adjust the variance during sampling. Following the elimination of variance, the reverse process of the diffusion model transforms into a deterministic procedure which means that starting from a certain noise, only a unique microstructure can be achieved. Consequently, the adjustability of the DDIM sampling process not only governs the randomness of microstructure generation through variance manipulation but also allows for the exploitation of deterministic sampling to comprehensively explore the latent space of the diffusion model. This dual capability presents a promising avenue for controlling microstructure characteristics.

\subsubsection{Reconstruction considering manufacture deviation based on DDIM}
As demonstrated above, diffusion models have the capacity to learn both compact and diverse microstructures, expanding the design space for material optimization. This, in turn, significantly enhances the performance of microstructure design optimization \cite{2022arXiv220210558C}.

In practical applications, additive manufacturing technology is often employed for the realization of such microstructures. However, the inherent complexities and uncertainties associated with the manufacturing process introduce unpredictability. Factors like the precision of 3D printing and equipment wear can influence the outcome of the manufacturing process, potentially deviating from the expected performance of the theoretically optimal microstructure. Hence, it becomes imperative to investigate the impact of random defects that may arise during the manufacturing process on the mechanical behavior of materials \cite{RN682, RN636}.

DDIM offers an inherent means to quantitatively assess the uncertainty stemming from manufacture deviation. As evident from the discussion in section 2.2, when $\eta=0$, DDIM's variance setting aligns with that of DDPM, and every step of denoising in the reverse process contains randomness. However, when $\eta=0$, no random noise is present, resulting in deterministic sampling without manufacturing deviation. The value of $\eta$ is intricately tied to the degree of randomness in microstructure reconstruction, essentially indicating the extent of deviation from deterministic reconstruction. 

\begin{figure}[htbp]
	\centering
	\begin{subfigure}{0.2\linewidth}
		\centering
		\includegraphics[width=2.48cm,height=2.48cm]{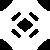}
		\caption{Original}
		\label{10.1}%文中引用该图片代号
	\end{subfigure}
	\centering
	\begin{subfigure}{0.2\linewidth}
		\centering
		\includegraphics[width=1.0\linewidth]{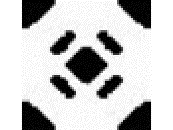}
		\caption{$\eta =0$}
		\label{10.2}%文中引用该图片代号
	\end{subfigure}
	\centering
	\begin{subfigure}{0.2\linewidth}
		\centering
		\includegraphics[width=1.0\linewidth]{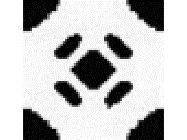}
		\caption{$\eta =0.1$}
		\label{10.3}%文中引用该图片代号
	\end{subfigure}
    \centering
	\begin{subfigure}{0.2\linewidth}
		\centering
		\includegraphics[width=1.0\linewidth]{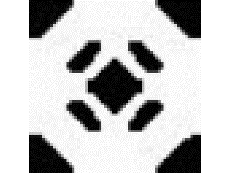}
		\caption{$\eta =0.2$}
		\label{10.4}%文中引用该图片代号
	\end{subfigure}

    \centering
	\begin{subfigure}{0.2\linewidth}
		\centering
		\includegraphics[width=1.0\linewidth]{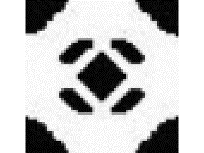}
		\caption{$\eta =0.4$}
		\label{10.5}%文中引用该图片代号
	\end{subfigure}
    \centering
	\begin{subfigure}{0.2\linewidth}
		\centering
		\includegraphics[width=1.0\linewidth]{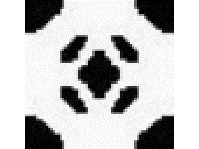}
		\caption{$\eta =0.6$}
		\label{10.6}%文中引用该图片代号
	\end{subfigure}
    \centering
	\begin{subfigure}{0.2\linewidth}
		\centering
		\includegraphics[width=1.0\linewidth]{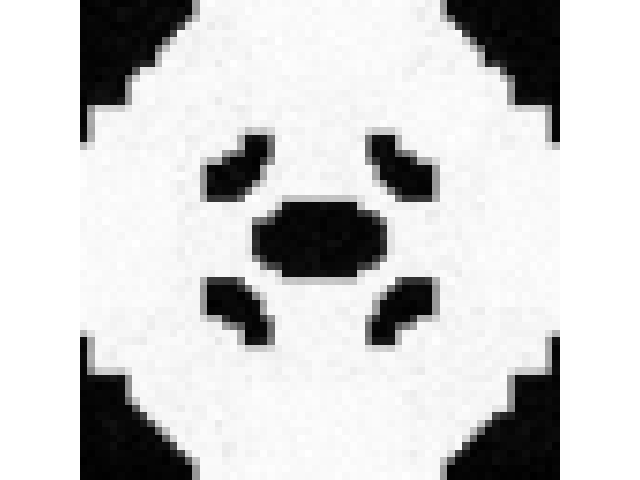}
		\caption{$\eta =0.8$}
		\label{10.7}%文中引用该图片代号
	\end{subfigure}
    \centering
	\begin{subfigure}{0.2\linewidth}
		\centering
		\includegraphics[width=1.0\linewidth]{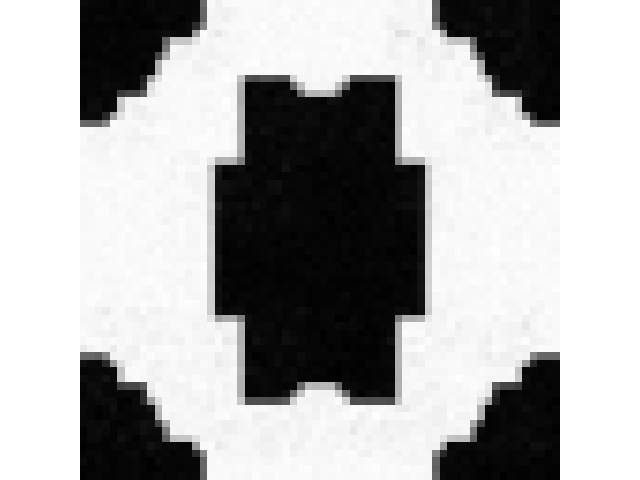}
		\caption{$\eta =1.0$}
		\label{10.8}%文中引用该图片代号
	\end{subfigure}
    \caption{Microstructure reconstruction considering manufacturing deviation}
	\label{label10}
\end{figure}

As depicted in Figure ~\ref*{label10}, this concept is illustrated by introducing varying levels of randomness in metamaterial reconstruction. As the parameter $\eta$ increases, the degree of deviation in manufacturing is also gradually increasing. Specifically, the four black rectangles located on the diagonal gradually deviate from their regular shapes, adopting irregular forms. A similar effect can be observed in the black square positioned at the center of the microstructure.

Randomness in diffusion model is the guarantee of diversity. Conversely, removing this randomness from diffusion models yields a deterministic path for microstructure generation. Nevertheless, during real-world manufacturing processes, errors are inevitable, and the degree of control over randomness within diffusion model precisely offers a means to quantify uncertainty in the actual formation of microstructures.

\subsubsection{Randomness control and gradient material generation based on DDIM}
The diffusion model can be seen as a multi-layer VAE. Similar to VAE, the diffusion model encodes raw data into a latent space and subsequently utilizes a decoder to map the samples from the latent space back to the original data space. However, the key distinction lies in the fact that the diffusion model doesn't involve dimensionality reduction operations. Instead, it establishes an equal-dimensional mapping between the latent space distribution and the original data distribution, where the latent space represents the noise distribution. Furthermore, since the diffusion model transforms discrete original image data into a continuous manifold structure, the mapping described above is inherently continuous as well. Building on this foundation, the diffusion model is capable of achieving the dynamic evolution of two microstructures by continuously interpolating between two noise distributions. This capability not only allows for the control of microstructure randomness but also offers a quantifiable design space for the inverse design of random materials through distance measurements in latent spatial interpolations \cite{RN642}.

It is worth noting that DDIM is sensitive to noise distribution. When linear interpolation is employed, $\lambda z_1+\left( 1-\lambda \right) z_2$ deviates from a normal distribution due to the superposition of normal distributions. Therefore, in this study, spherical interpolation was adopted \cite{Shoemake1985AnimatingRW, 2020arXiv201002502S}, which can be expressed as follows:
\begin{equation}
    \begin{array}{c}
        \mathbf{x}_{T}^{\left( \alpha \right)}=\frac{\sin \left( \left( 1-\alpha \right) \theta \right)}{\sin \left( \theta \right)}\mathbf{z}_{T}^{\left( 0 \right)}+\frac{\sin \left( \alpha \theta \right)}{\sin \left( \theta \right)}\mathbf{z}_{T}^{\left( 1 \right)},
    \end{array}
\end{equation}
where $\theta =\arccos \left( \frac{\left( \mathbf{z}_{T}^{\left( 0 \right)} \right) ^{\text{T}}\mathbf{z}_{T}^{\left( 1 \right)}}{||\mathbf{z}_{T}^{\left( 0 \right)}||\mathbf{z}_{T}^{\left( 1 \right)}||} \right)$, and $\mathbf{z}$ follows standard normal distribution. 

\begin{figure}[htbp]
	\centering
	\begin{subfigure}{1\linewidth}
		\centering
		\includegraphics[width=0.95\linewidth]{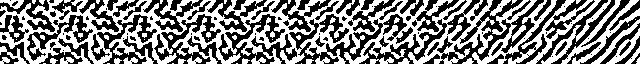}
		\caption{Random materials to directed random materials}
		\label{label11.1}%文中引用该图片代号
	\end{subfigure}

	\centering
	\begin{subfigure}{1\linewidth}
		\centering
		\includegraphics[width=0.95\linewidth]{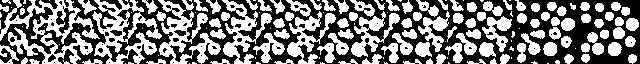}
		\caption{Random material to circular inclusion material}
		\label{label11.2}%文中引用该图片代号
	\end{subfigure}

	\centering
	\begin{subfigure}{1\linewidth}
		\centering
		\includegraphics[width=0.95\linewidth]{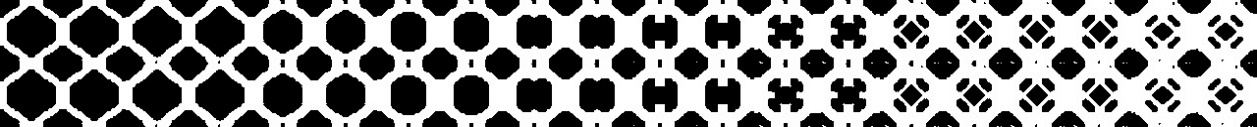}
		\caption{Generated gradient material}
		\label{label11.3}%文中引用该图片代号
	\end{subfigure}
    \caption{Randomness control and gradient materials based on DDIM}
	\label{label11}
\end{figure}

Figure ~\ref*{label11.1} and ~\ref*{label11.2} illustrate the continuous interpolation decoding in the latent spaces of two microstructures, achieving dynamic evolution from random materials to texture materials and from random materials to circular inclusion materials, respectively. This method offers the potential to characterize and regulate the randomness within microstructures. The interpolation distance between different microstructures in the latent space serves as a measure of randomness, and various levels of randomness can be controlled in the microstructure by adjusting the distance from a state of complete randomness. This also provides a foundation for the controllable application of randomness in material design. 

Using the method outlined above, it also becomes possible to create gradient structures without the need for direct generation. Instead, gradient structures can be generated by combining intermediate structures from an extensive dynamic evolution process. Figure ~\ref*{label11.3} generates gradient metamaterials based on continuity decoding in latent space.

\subsection{Conditional generation of 3D random materials based on DDPM}
\begin{figure}[htbp]
	\centering
	\begin{subfigure}{0.3\linewidth}
		\centering
		\includegraphics[width=1.0\linewidth]{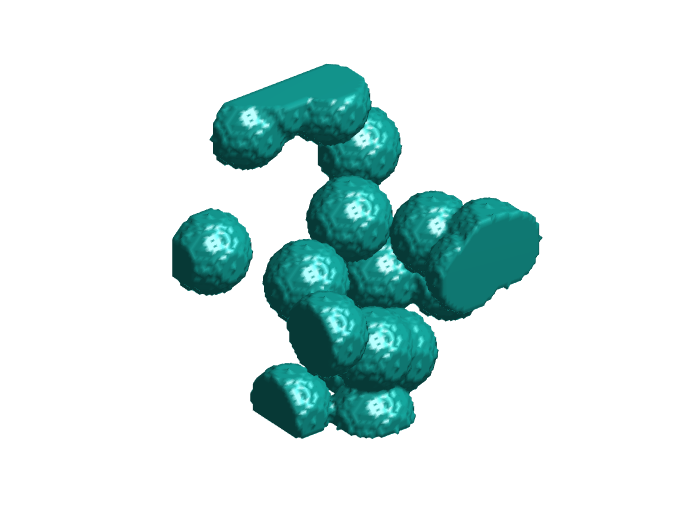}
		\caption{}
		\label{label12.4}%文中引用该图片代号
	\end{subfigure}
	\centering
	\begin{subfigure}{0.3\linewidth}
		\centering
		\includegraphics[width=1.0\linewidth]{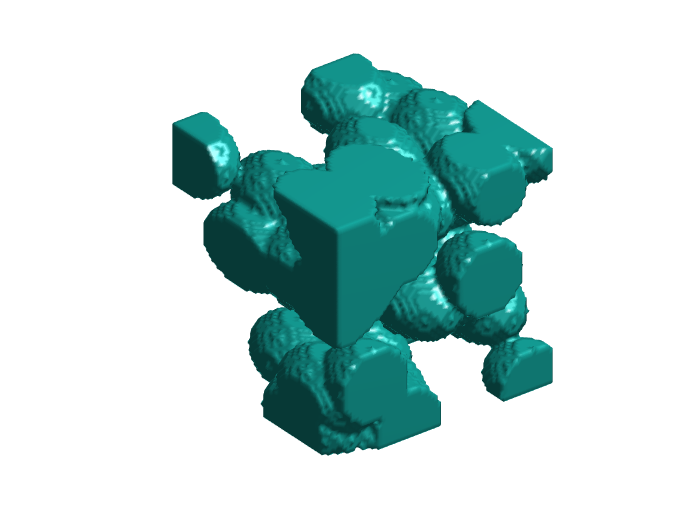}
		\caption{}
		\label{label12.5}%文中引用该图片代号
	\end{subfigure}
	\centering
	\begin{subfigure}{0.3\linewidth}
		\centering
		\includegraphics[width=1.0\linewidth]{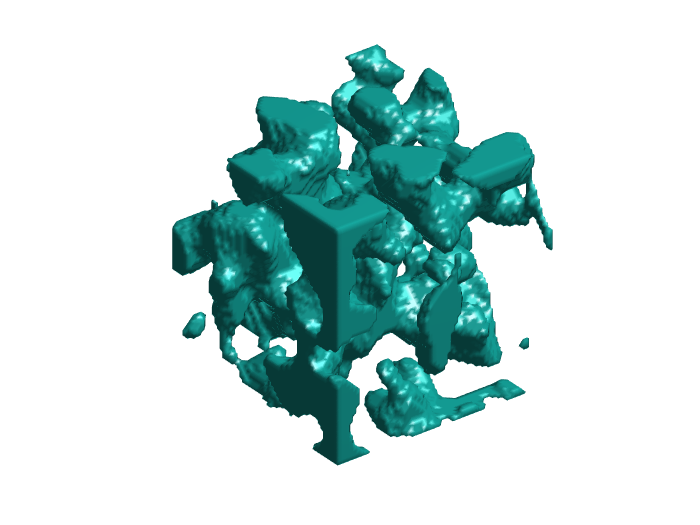}
		\caption{}
		\label{label12.6}%文中引用该图片代号
	\end{subfigure}

	\centering
	\begin{subfigure}{0.3\linewidth}
		\centering
		\includegraphics[width=1.0\linewidth]{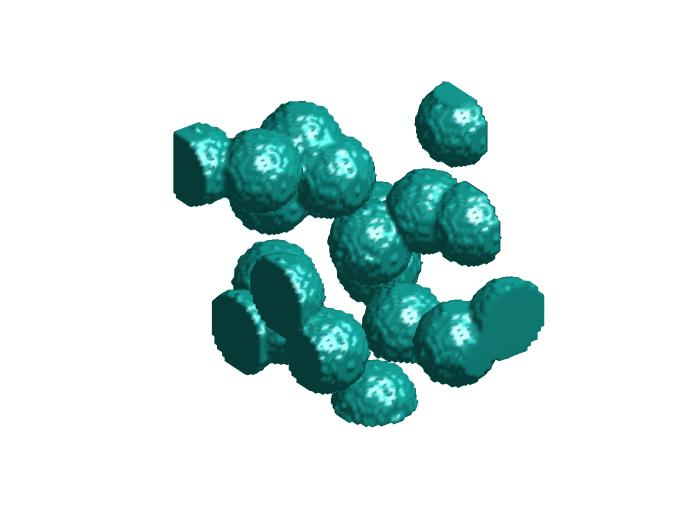}
		\caption{}
		\label{label12.1}%文中引用该图片代号
	\end{subfigure}
	\centering
	\begin{subfigure}{0.3\linewidth}
		\centering
		\includegraphics[width=1.0\linewidth]{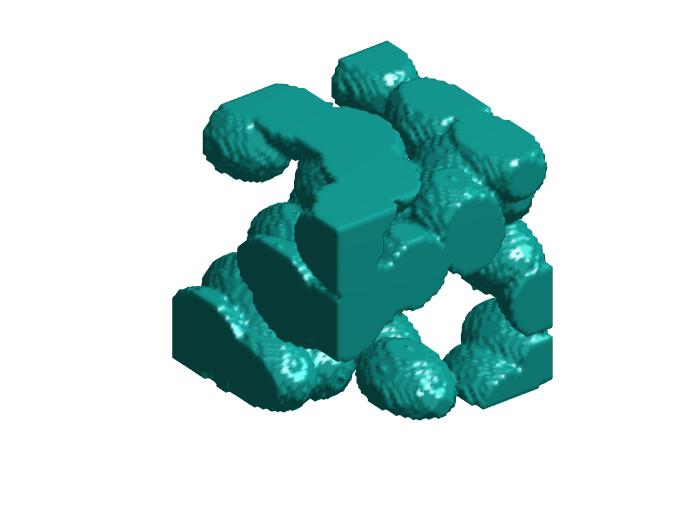}
		\caption{}
		\label{label12.2}%文中引用该图片代号
	\end{subfigure}
	\centering
	\begin{subfigure}{0.3\linewidth}
		\centering
		\includegraphics[width=1.0\linewidth]{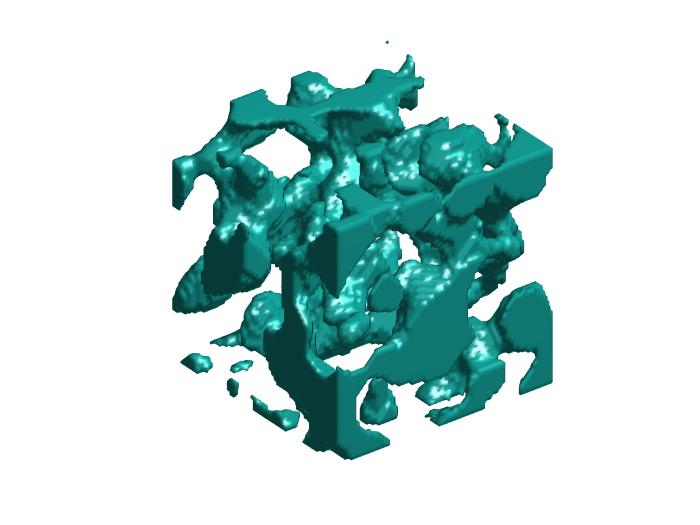}
		\caption{}
		\label{label12.3}%文中引用该图片代号
	\end{subfigure}
    \caption{Comparison between original three-dimensional microstructure and reconstructed microstructure of composite materials; (a):Original spherical inclusion; (b):Original ellipsoidal inclusion; (c):Original random porous material; (d):Generated spherical inclusion; (e):Generated ellipsoidal inclusion; (f):Generated random porous material}
	\label{label12}
\end{figure}

In microstructure reconstruction of composite, this study extends the diffusion model framework to 3D, and successfully achieves 3D reconstruction of spherical inclusion materials, ellipsoidal inclusion materials, and random materials, all with sizes of $64\times64\times64$. The comparison between the original three-dimensional microstructure and the generated three-dimensional microstructure is shown in the Figure ~\ref*{label12}. 

\begin{figure}[htbp]
	\centering
	\begin{subfigure}{0.3\linewidth}
		\centering
		\includegraphics[width=1.0\linewidth]{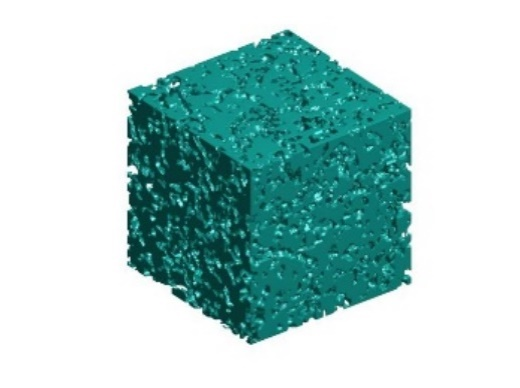}
		\caption{}
		\label{label13.1}%文中引用该图片代号
	\end{subfigure}
	\centering
	\begin{subfigure}{0.3\linewidth}
		\centering
		\includegraphics[width=1.0\linewidth]{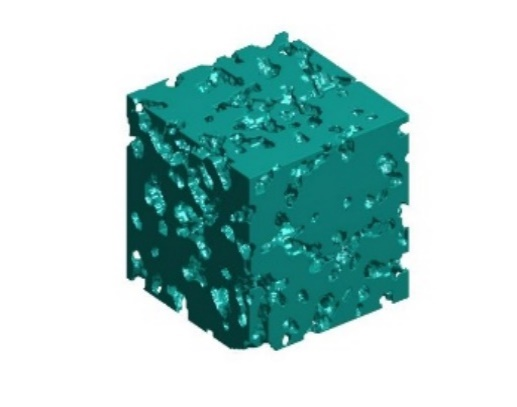}
		\caption{}
		\label{label13.2}%文中引用该图片代号
	\end{subfigure}
	\centering
	\begin{subfigure}{0.3\linewidth}
		\centering
		\includegraphics[width=1.0\linewidth]{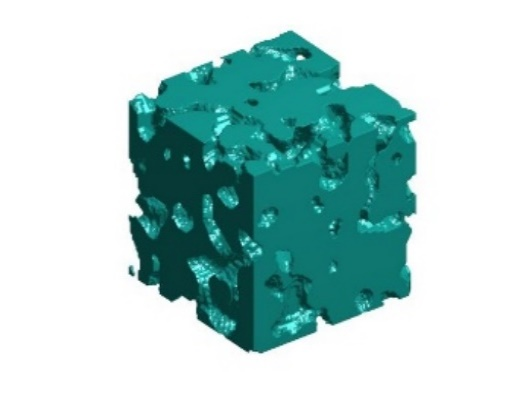}
		\caption{}
		\label{label13.3}%文中引用该图片代号
	\end{subfigure}

    \centering
	\begin{subfigure}{0.3\linewidth}
		\centering
		\includegraphics[width=1.0\linewidth]{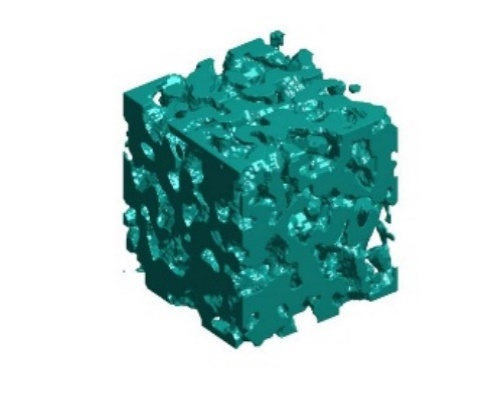}
		\caption{}
		\label{label13.4}%文中引用该图片代号
	\end{subfigure}
	\centering
	\begin{subfigure}{0.3\linewidth}
		\centering
		\includegraphics[width=1.0\linewidth]{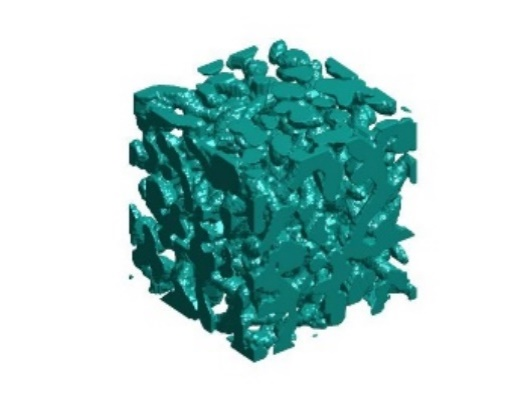}
		\caption{}
		\label{label13.5}%文中引用该图片代号
	\end{subfigure}
	\centering
	\begin{subfigure}{0.3\linewidth}
		\centering
		\includegraphics[width=1.0\linewidth]{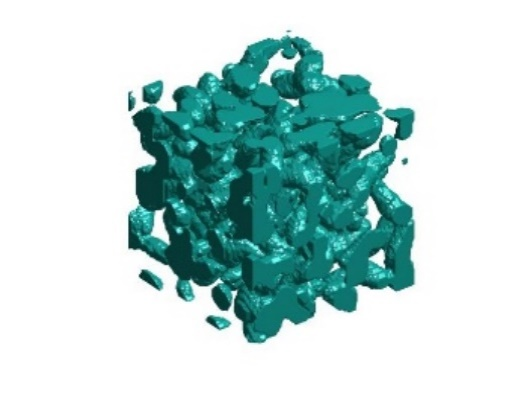}
		\caption{}
		\label{label13.6}%文中引用该图片代号
	\end{subfigure}
    \caption{Reconstruction of microstructure of three-dimensional composite materials;(a)permeability=0.14; (b)permeability=0.44; (c)permeability=0.66; (d)permeability=1.83; (e)permeability=4.57; (f)permeability=10.93}
	\label{label13}
\end{figure}

Material performance is intricately tied to its three-dimensional microstructure. This connection allows for the intentional design of microstructures to achieve desired performance objectives. In this study, the permeability of random porous materials serves as an illustrative example. Leveraging the mapping relationship between microstructure and permeability, conditional generation was performed on three-dimensional random materials within a specific permeability range. Specifically, the permeability of three-dimensional materials is divided into six ranges as number label, each of which serves as a feature encoding embedding combined with image and time encoding. The framework of the network is shown in the section 3.1. The conditional diffusion model employed in this study has proven effective in generating three-dimensional microstructures with permeability falling within six specific ranges: (0, 0.2), (0.2, 0.5), (0.5, 1.5), (1.5, 3.0), (3.0, 5.0), and (5.0, 5.0+). The specific results are shown in Figure ~\ref*{label13}.

The verification of the permeability of random porous materials is carried out through the lattice Boltzmann numerical method,
\begin{equation}
	f_i\left( \mathbf{x}+\mathbf{c}_i\delta t,t+\delta t \right) -f_i\left( \mathbf{x,}t \right) =-\tau ^{-1}\cdot \left( f_i\left( \mathbf{x,}t \right) -f_{i\left( eq \right)}\left( \mathbf{x,}t \right) \right),
\end{equation}
where $f$ is the particle velocity distribution function, and the Boltzmann equation is essentially a conservative description of the spatiotemporal changes of $f$. $\mathbf{x}$ and $\mathbf{c}$ are the positions and velocities of particles, respectively. $\tau$ is the relaxation time. $f_{i\left( eq \right)}$ is the equilibrium distribution function. In this study, slip-free and rebound boundary conditions are applied at the two-phase interface. The initial fluid velocity is set to zero, and flow is induced by a constant pressure difference in the structure along the transport direction \cite{10.1063/1.869307}. The initial condition entails a linear pressure gradient, and the relaxation time is maintained at 1.0. 

Permeability can be determined using Darcy's law, as follows:
\begin{equation}
    \bar{u}=-\frac{\kappa \Delta p}{\mu d}.
\end{equation}

\begin{figure}[htbp]
	\centering
	\begin{subfigure}{0.3\linewidth}
		\centering
		\includegraphics[width=1.0\linewidth]{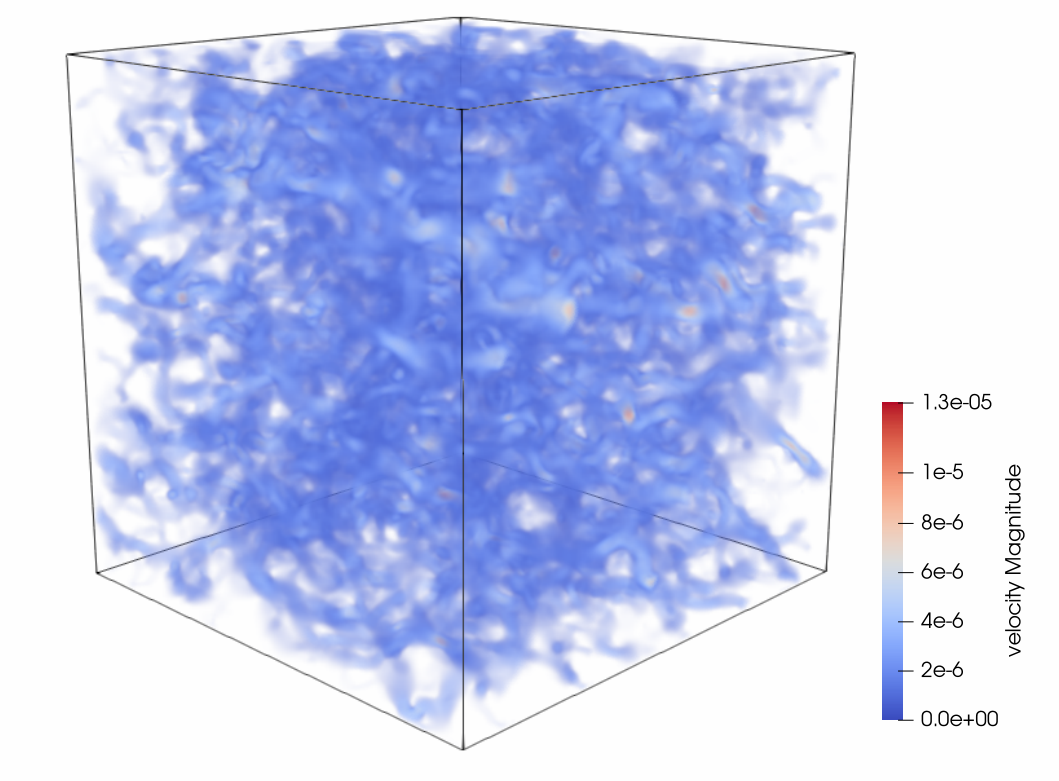}
		\caption{}
		\label{label14.1}%文中引用该图片代号
	\end{subfigure}
	\centering
	\begin{subfigure}{0.3\linewidth}
		\centering
		\includegraphics[width=1.0\linewidth]{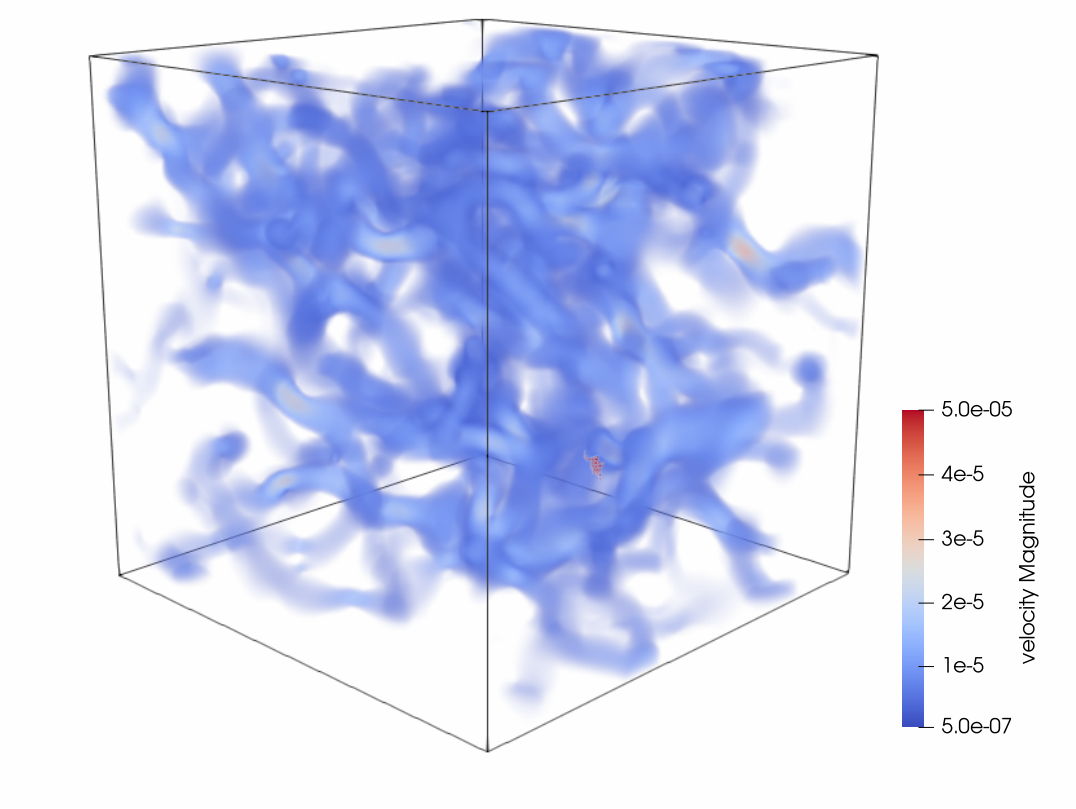}
		\caption{}
		\label{label14.2}%文中引用该图片代号
	\end{subfigure}
	\centering
	\begin{subfigure}{0.3\linewidth}
		\centering
		\includegraphics[width=1.0\linewidth]{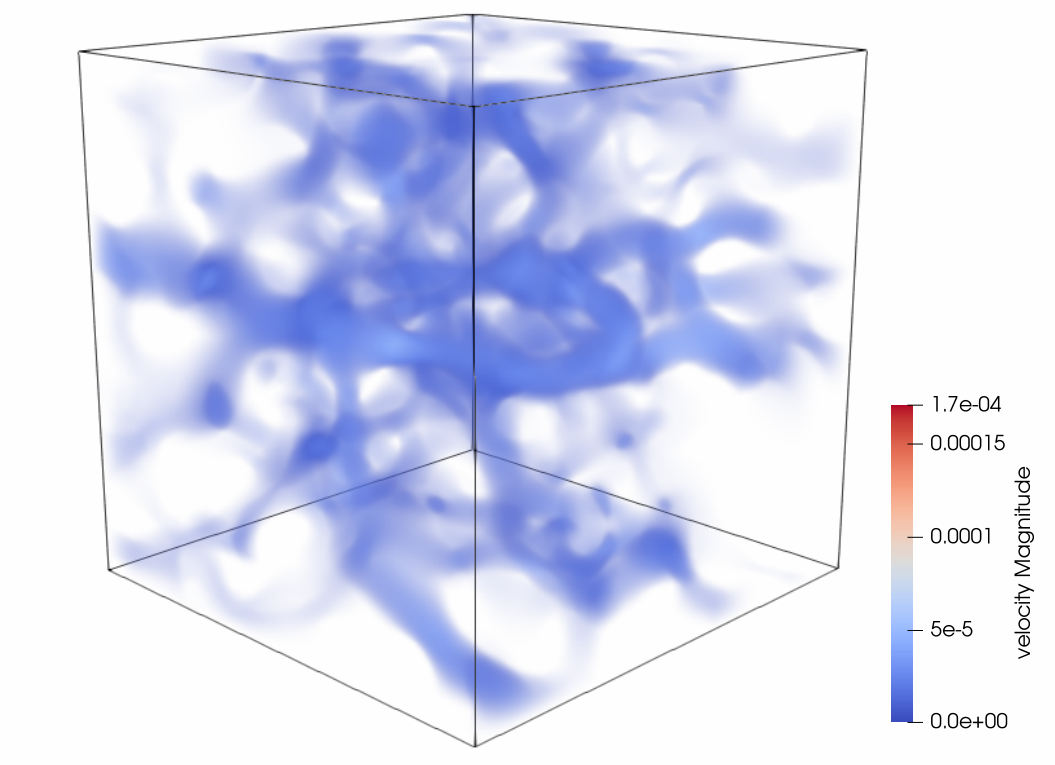}
		\caption{}
		\label{label14.3}%文中引用该图片代号
	\end{subfigure}

    \centering
	\begin{subfigure}{0.3\linewidth}
		\centering
		\includegraphics[width=1.0\linewidth]{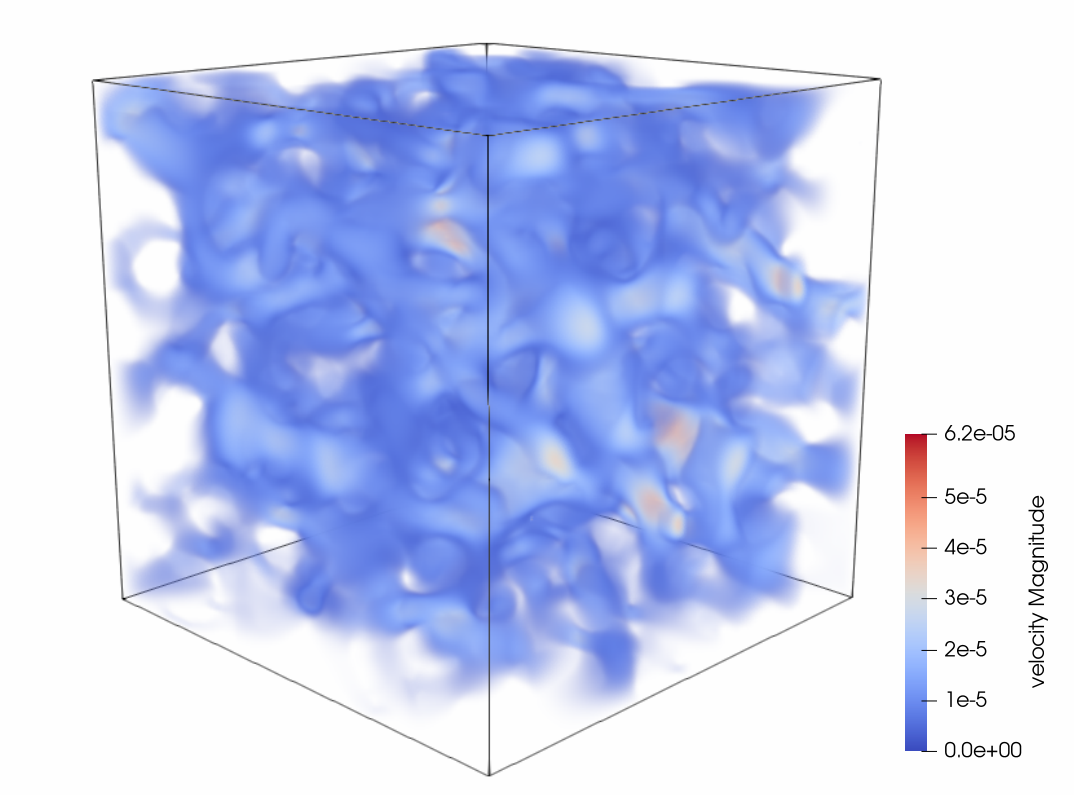}
		\caption{}
		\label{label14.4}%文中引用该图片代号
	\end{subfigure}
	\centering
	\begin{subfigure}{0.3\linewidth}
		\centering
		\includegraphics[width=1.0\linewidth]{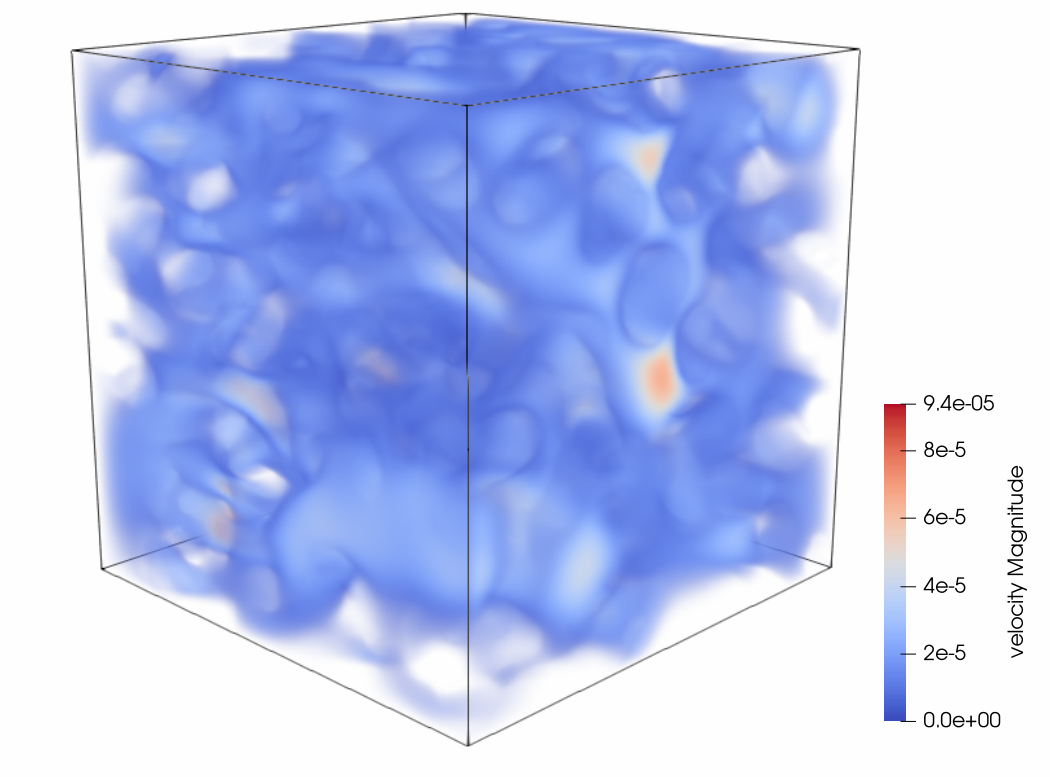}
		\caption{}
		\label{label14.5}%文中引用该图片代号
	\end{subfigure}
	\centering
	\begin{subfigure}{0.3\linewidth}
		\centering
		\includegraphics[width=1.0\linewidth]{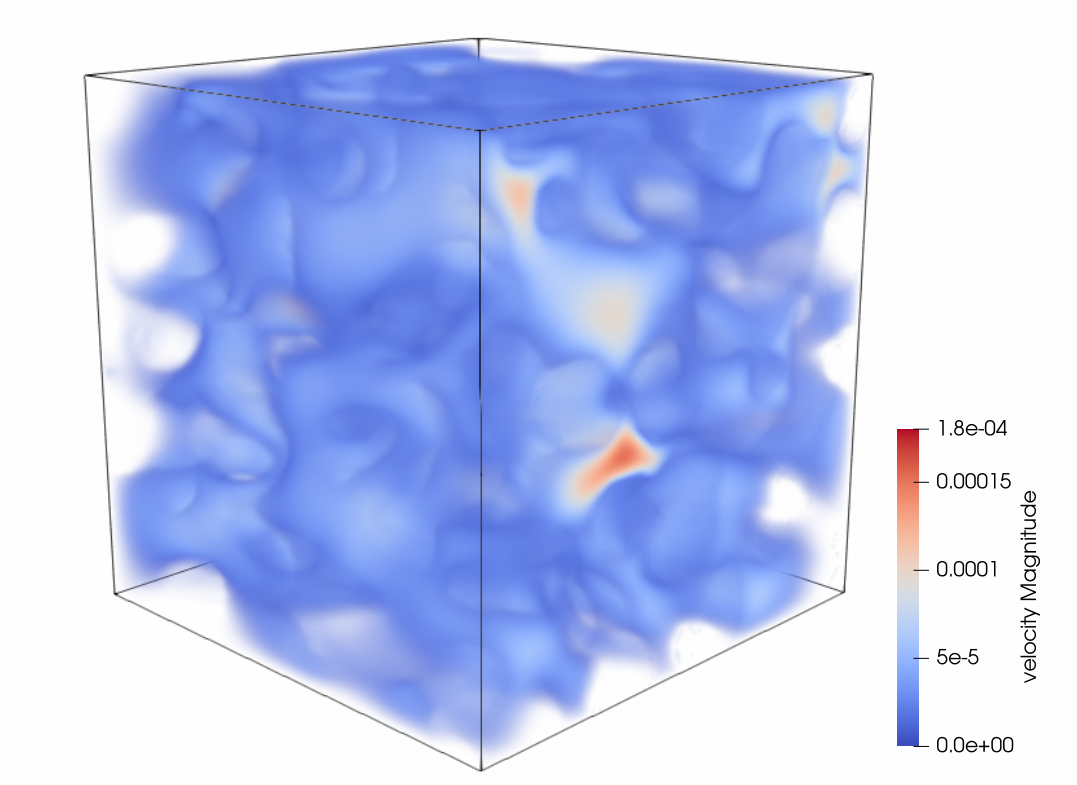}
		\caption{}
		\label{label14.6}%文中引用该图片代号
	\end{subfigure}
    \caption{Velocity distribution in generated random porous material;(a)permeability=0.14; (b)permeability=0.44; (c)permeability=0.66; (d)permeability=1.83; (e)permeability=4.57; (f)permeability=10.93}
	\label{label14}
\end{figure}

Among them, $\bar{u}$ is the average velocity, $\mu$ is the fluid dynamic viscosity, and $\varDelta p/d$ is the pressure gradient. Under the above conditions, the velocity distribution of the fluid in the pores inside the random porous material is shown in the Figure ~\ref*{label14}. The permeability of random materials is 0.14, 0.44, 0.66, 1.83, 4.57, and 10.93, respectively, which follows the range represented by feature encoding embedding.

\section{Conclusions}
In this study, DDPM was successfully used to reconstruct the two-dimensional microstructure of various composite materials, including particle inclusion materials, quasi random materials, random materials, etc. In order to evaluate the quality of reconstruction, statistical comparisons were made using descriptors such as the two-point correlation function, linear path function, and Fourier descriptors, which quantitatively compared spatial relationships and boundary shapes. The results demonstrate the morphological similarity between the generated microstructure and the original microstructure, highlighting the diffusion model's efficacy in reconstruction. Based on the above foundation, the study also utilized DDIM for continuous interpolation in the latent space, enabling the regulation of microstructure randomness and the creation of gradient materials. This provides a new approach for the controllable application of randomness in material design.

Moreover, this study extended two-dimensional microstructure reconstruction to a three-dimensional framework and successfully reconstructed the three-dimensional microstructure of spherical inclusions, ellipsoidal inclusions, and random porous materials. Additionally, permeability was incorporated as the feature encoding embedding, allowing for the conditional generation of three-dimensional microstructures within a defined permeability range, providing a feasible approach for inverse design of three-dimensional random materials.

However, it's important to acknowledge that this work represents a preliminary attempt in material inverse generation based on diffusion model and is limited to defined permeability ranges. Future research should explore precise permeability control and microstructure design under multi-physical field coupling. Additionally, while diffusion models demonstrate excellence in microstructure reconstruction, they present challenges when used for forward optimization in material design because of their extensive latent spatial dimensions and unclear semantics. Combining the latent spatial dimensionality reduction operation of VAE and the excellent generation ability of diffusion models is also a direction that needs to be worked on in the future.

\section*{Declaration of competing interest}
The authors declare that they have no known competing financial interests or personal relationships that could have appeared to influence the work reported in this paper.

\section*{Data Availability Statement}
Data will be made available on request.

\section*{Acknowledgements}
The financial support provided by National Natural Science Foundation of China (Grant No. 52078361) is greatly appreciated.

% \begin{itemize}
% \item document style
% \item baselineskip
% \item front matter
% \item keywords and MSC codes
% \item theorems, definitions and proofs
% \item lables of enumerations
% \item citation style and labeling.
% \end{itemize}

% \section{Front matter}

% The author names and affiliations could be formatted in two ways:
% \begin{enumerate}[(1)]
% \item Group the authors per affiliation.
% \item Use footnotes to indicate the affiliations.
% \end{enumerate}
% See the front matter of this document for examples. You are recommended to conform your choice to the journal you are submitting to.

% \section{Bibliography styles}

% There are various bibliography styles available. You can select the style of your choice in the preamble of this document. These styles are Elsevier styles based on standard styles like Harvard and Vancouver. Please use Bib\TeX\ to generate your bibliography and include DOIs whenever available.

% Here are two sample references: .

% \section*{References}

\bibliography{mybibfile}

\end{document}